\documentclass[12pt,reqno]{amsart}
\usepackage{amssymb}
\usepackage{geometry}
\usepackage[italian,english]{babel}
\usepackage[utf8]{inputenc}
\usepackage{amsmath, amsfonts, amsthm}
\usepackage{graphicx}
\usepackage{pgfplots}
\usepackage{paralist}
\usepackage{mathtools, mathabx,bigints,cancel}
\numberwithin{equation}{section}
\usepackage{diagbox}
\usepackage{booktabs} 
\usepackage{comment} 
\usepackage{enumitem}
\usepackage{subfig}
\usepackage{bm}
\usepackage{url}
\usepackage{lineno}
\usepackage{hyperref}
\usepackage{cleveref}
\usepackage{siunitx}
\usepackage{longtable}
\usepackage{array}
\newtheorem{theorem}{Theorem}[section]
\newtheorem{lemma}[theorem]{Lemma}
\newtheorem{proposition}[theorem]{Proposition}

\newtheorem{definition}[theorem]{Definition}

\newtheorem{remark}[theorem]{Remark}

\newtheorem{notation}[theorem]{Notation}
\theoremstyle{definition}

\newcommand{\N}{\mathbf{N}}
\newcommand{\dive}{\mathrm{div}}

\newcommand{\R}{\mathbb{R}}

\newcommand{\norm}[1]{\left\Vert#1\right\Vert}
\newcommand{\abs}[1]{\left\vert#1\right\vert}
\usepackage{marginnote}

\begin{document}

\title
[Monotonicity Principle and \lq\lq $p$-Laplace Signature\rq\rq]
{Monotonicity Principle and \lq\lq $p$-Laplace Signature\rq\rq \ for Tomography in Nonlinear Elliptic Inverse Problems}

\author[Gianpaolo Piscitelli, Vincenzo Mottola, Antonello Tamburrino]{Gianpaolo Piscitelli$^1$, Vincenzo Mottola$^{2,3}$, Antonello Tamburrino$^{2,3}$}

\footnotetext[1]{Dipartimento di Scienze Economiche Giuridiche Informatiche e Motorie, Universit\`a degli Studi di Napoli Parthenope, Via Guglielmo Pepe, Rione Gescal, 80035 Nola (NA), Italy.\\
Email: {\rm gianpaolo.piscitelli@uniparthenope.it}.}

\footnotetext[2]{Dipartimento di Ingegneria Elettrica e dell'Informazione \lq\lq M. Scarano\rq\rq, Universit\`a degli Studi di Cassino e del Lazio Meridionale, Via G. Di Biasio n. 43, 03043 Cassino (FR), Italy.\\
Email: {\textrm vincenzo.mottola@unicas.it  (corresponding author), antonello.tamburrino@unicas.it}.}

\footnotetext[3]{EUT+ Institute of Nanomaterials and Nanotechnologies-EUTINN, European University of Technology, European Union.}

\setcounter{tocdepth}{2}

\begin{abstract}

This paper proposes a framework for treating the inverse obstacle problem for nonlinear elliptic equations with nonlinear materials.

The problem is challenging because nonlinear materials exhibit a rich diversity of scenarios to consider, since nonlinearity can take different forms. In this article, after categorizing the nonlinearities into a few fundamental classes, a dedicated imaging method is proposed for each class, derived by combining two powerful concepts: the Monotonicity Principle (MP) and the $p-$Laplace Signature (pLS).


The Monotonicity Principle (MP), recently extended to nonlinear materials, provides a monotonic relationship between the material property and the measured quantity (the Average Dirichlet-to-Neumann map) that can be \lq\lq inverted\rq\rq \ to find the shape of anomalies.

The $p-$Laplace Signature (pLS) allows for modelling the solution of an elliptic PDE with nonlinear materials, for large or small boundary data, in terms of a proper $p-$Laplace equation that captures the essence (the signature) of the problem. For example, pLS with $p=2$ allows the reduction of a nonlinear elliptic PDE to a linear one, providing a powerful bridge for applying imaging methods and algorithms developed for linear materials.

In this contribution, the two pillars of MP and pLS are combined in new imaging methods to enlarge the class of nonlinearity that can be treated within the inverse obstacle problem. Moreover, the theoretical limits of the methods are provided in the ideal case of noise-free measurements: outer-support reconstruction when $p=2$, and convex-hull reconstruction when $p\neq2$. 



\end{abstract}

\maketitle

\section{Introduction}
This work falls in the framework of the nonlinear generalization of the classical Calderón problem 
\cite{alessandrini1989remark,calderon1980inverse,calderon2006inverse}, 
focusing on the analysis of nonlinear elliptic equations of the form
\begin{equation}
\label{problem}
\begin{cases}
\dive\Big(\gamma (x, |\nabla u(x)|) \nabla u (x)\Big) =0 \quad &\text{in }\Omega, \vspace{0.2cm}\\
u(x) =f(x) &\text{on }\partial\Omega,
\end{cases}
\end{equation}
where $f$ is the applied boundary datum, $u$ is the solution, and 
$\Omega\subset\R^n$, $n \geq 2,$ is a bounded Lipschitz domain representing the region occupied by the material.  

Problem \eqref{problem} naturally models several physical scenarios. For instance, with reference to electromagnetism, problem \eqref{problem} models (i) nonlinear steady-state conduction, (ii) magnetostatic, and (iii) electrostatic. The first case is at the foundation of \emph{Electrical Resistance Tomography} (ERT) where $\gamma=\sigma$ is the electrical conductivity, the second case is at the foundation of \emph{Magnetic Inductance Tomography} (MIT) where $\gamma=\mu$ is the magnetic permeability, and the third case is at the foundation of \emph{Electrical Capacitance Tomography} (ECT) where $\gamma=\varepsilon$ is the dielectric permittivity.

Since nonlinear materials are widely used in various engineering fields, the development of tomographic methods capable of addressing nonlinearities is highly desirable. 

Prominent examples of real-world problems involving nonlinear magnetic materials include the non-destructive evaluation of steel pipelines used in steam generators \cite{shi2015theory} of nuclear power plants and the non-destructive testing of railway tracks \cite{MORDIA2025163}. Beyond industrial diagnostics, other applications include reconstructing the shape of metallic and potentially magnetic objects inside closed containers for security screening \cite{book:Dorn18}. Similarly, in civil engineering, concrete inspection often requires monitoring of steel reinforcement bars; these rebars can exhibit nonlinear magnetic behavior and are highly susceptible to corrosion, thus demanding advanced non-destructive monitoring techniques \cite{art:So05}.

Regarding steady-state currents, a representative example of nonlinear materials is the class of superconducting materials, which exhibit an electrical conductivity that decreases as a function of the electric field \cite{rhyner1993magnetic}. These materials are widely employed in several key technologies, such as energy storage systems and superconducting magnets \cite{seidel2015applied}. Furthermore, materials with nonlinear electrical conductivities are intentionally engineered for electric-field-grading applications \cite{boucher2018interest}. Finally, human tissues also deserve attention, as they have been shown to exhibit nonlinear electrical behavior under specific conditions \cite{301734,corovic2013modeling}.

In addition to magnetic and conductive materials, nonlinear behaviour is observed in dielectric materials. A prime class of such materials is ferroelectric, such as barium titanate ($BaTiO_3$) \cite{PhysRevB.85.224111}. In these systems, nonlinearity is widely exploited in tunable microwave devices, non-volatile memories, and high-density energy storage capacitors. Another significant nonlinear dielectric response occurs within semiconductor structures, most notably in Schottky barriers and $p$-$n$ junctions \cite{art:diode_nl}. 

Moreover, the growing interest in the design and fabrication of advanced composite materials and nonlinear electromagnetic metamaterials \cite{Poutrina_2010} has further motivated the development of imaging methods for nonlinear media. As these materials become increasingly adopted in a wide range of applications, the demand for reliable non-destructive testing and imaging techniques is expected to grow significantly.

\begin{figure}[ht]
    \centering
\includegraphics[width=0.6\textwidth]{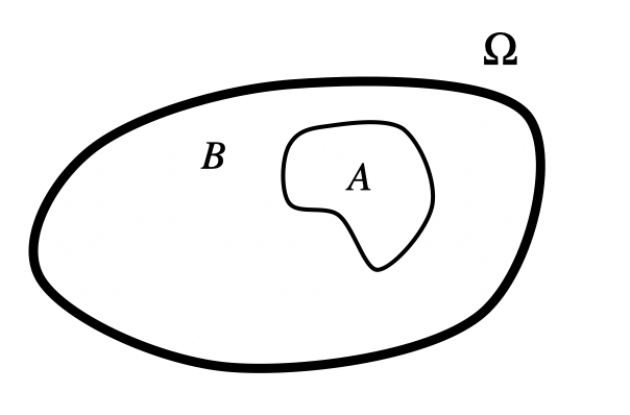}
    \caption{A two-phase configuration. The anomaly $A$ has an arbitrary shape and topology. It may be made up of one or more connected components. The measured data are gathered on the boundary $\partial \Omega$ of the domain of interest. The picture is in \cite[Figure 1]{corboesposito2024thep0laplacesignature}, and it is courtesy of the Siam Journal of Imaging Science. }
    \label{fig_01_BmenoA}
\end{figure}

In this contribution, the nonlinear inverse obstacle problem for elliptic PDEs of the form \eqref{problem} is addressed. Specifically, the target is to retrieve from boundary measurements the shape of an anomaly $A$ embedded in a background region $\Omega$, as shown in Figure~\ref{fig_01_BmenoA}. The constitutive relationship for the material contained in the anomaly and/or in the background is nonlinear. The anomaly $A$ has an arbitrary shape and topology and may consist of multiple connected components. In any case, any connected component of $A$ is a bounded Lipschitz domain.

To date, there are three main approaches that address the problem's physical-mathematical structure of the inverse obstacle problem: $p-$Laplace Monotonicity Principle Method (P-MPM), $p-$Laplace Enclosure Method (P-EM), and the full nonlinear Monotonicity Principle based on the Average Dirichlet-to-Neumann operator (A-MPM). P-MPM treats specific nonlinear constitutive relationships \cite{Salo2012_IP,brander2016calderon,brander2018superconductive,guo2016inverse,brander2018monotonicity,hauer2015p}, where the authors focused on the special (nonlinear) case of the {Calder\'on} problem modeled by the $p$-Laplace operator and closer variants. P-EM focuses on the Enclosure Method extended to the {Calder\'on} problem modeled by the $p$-Laplace operator \cite{brander2015enclosure}.
Finally, the treatment of more general classes of nonlinearities has been achieved by means of the recently introduced \emph{Monotonicity Principle Method} (A-MPM) for nonlinear materials, where the concept of the average DtN (A-DtN) map replaces that of the classical DtN map \cite{corboesposito2021monotonicity,corboesposito2024piecewise,mottola2025theinverseobstacle,mottola2026corrigendumtheinverseobstacle,mottola2024imaging}.
In ideal conditions, both P-MPM and P-EM are capable of retrieving the convex hull of the anomaly $A$ but for the $p-$Laplace problem and closer variants, while A-MPM is capable of retrieving the outer support of a general nonlinear anomaly $A$, but embedded in a linear background. 

It is worth noting that both the P-MPM and P-EM can be naturally expressed in terms of the A-DtN operator, owing to the proportionality between the A-DtN and classical DtN operators for the $p$-Laplace problem (see, e.g., the proof of \Cref{thm_converse_p_esteso}). This highlights the central role of the A-DtN operator in providing a unified treatment of nonlinear variants of the Calderón problem.

Aside from A-MPM, another recent development capable of extending the class of nonlinearities that can be treated in the inverse obstacle problem is the so-called \emph{$p-$Laplace Signature} (pLS). pLS allows one to reduce a nonlinear elliptic PDE to a weighted $p-$Laplace problem in the limiting regimes of large or small boundary data \cite{corboesposito2024theplaplacesignature,corboesposito2024thep0laplacesignature}. Although pLS is not an imaging method, it provides a solid $p-$Laplacian description of the asymptotic behavior of nonlinear materials, which forms the basis for developing new imaging methods.

A-MPM and the pLS, the most general \lq\lq tools\rq\rq \ available, form the foundation of this contribution, which aims to enlarge the classes of nonlinearities that can be treated in the inverse obstacle problem. Moreover, A-MPM and pLS can be naturally combined together.

To appreciate the contribution of this paper, it is worth noting that the A-DtN and the MPM for nonlinear materials can be applied to a large class of non-linear materials \cite{corboesposito2021monotonicity,corboesposito2024piecewise}, enabling the reconstruction of upper and lower bounds for the anomaly $A$. A sharper result, the so-called Converse of MPM, proved in \cite{mottola2025theinverseobstacle} shows that the A-MPM reconstructs the outer support of $A$ when the background material is linear. In this work, this sharper result is extended by proving that (i) the A-MPM can provide the convex hull when the background is of $p$-Laplace type, and the anomaly is arbitrary, but with the growth exponent $q$ equal to $p$ (see \Cref{pLS_sec}), thus improving the result for P-MPM and P-EC, requiring the nonlinearity of the anomaly of same $p-$Laplace type. Subsequently, it is shown that combining the pLS with the A-MPM allows for (ii) the reconstruction of the outer support of $A$ even when the background material is nonlinear with $p = 2$ and the anomaly is made of an arbitrary nonlinear material, or (iii) the convex hull of $A$ for $p \neq 2$ and the anomaly has the growth exponent $q$ equal to $p$. From an abstract perspective, to date, the A-MPM allows generalising the type of nonlinearity for the anomaly from linear or $p$-Laplacian to a general nonlinear form, whereas the pLS enables generalising the background nonlinearity from $p$-Laplacian to an arbitrary one.

To obtain this, key results are \Cref{thm_converse_p_esteso}, proving that the A-MPM provides the convex hull when the background is of $p$-Laplace type, and 
\Cref{thm_conv_aDTN,thm_conv_aDTN0}, proving the convergence of the A-DtN map to the DtN map of the corresponding limiting $p$-Laplace problem, in both the small and large data regimes, enabling to prove that A-MPM can be extended to general nonlinear backgrounds (see \Cref{thm_inf_conv,thm_quasi_pls,thm_np_1,thm_np_2}), other than linear ($p=2$) or of $p-$Laplace type ($p \ne 2$).

The paper is organized as follows.
\Cref{sec_glimpse} provides a glimpse to the Monotonicity Principle and the $p-$Laplace Signature, together with the combined treatment proposed in this work.
\Cref{notation_sec} fixes the notation and describes the problem setting.
\Cref{sec_monotonicity_principle} states the main results from the Monotonicity Principle for nonlinear materials, and the $p-$Laplace Signature limiting behavior.
In \Cref{sec_p=2_quasi}, reconstruction methods based on the MP for bounded background properties are presented, detailing their theoretical limits and reconstruction strategies.
In \Cref{sec_p_methods}, the same results are presented in the case of a possibly unbounded or vanishing background material property.
Finally, \Cref{sec_conclusions} collects some conclusions and perspectives for future research.  


\section{A glimpse on MP and pLS}\label{sec_glimpse}
This Section gives a short introduction to the Monotonicity Principle for nonlinear materials (A-MPM), the Monotonicity Principle for $p-$Laplace and quasi $p-$Laplace materials (P-MPM), the Enclosure Method for $p-$Laplace materials (P-EM), the $p-$Laplace signature concepts (pLS), and the main results of this contribution, i.e., the extension of the existing A-MPM to broader classes of nonlinear inverse obstacle problems.

\subsection{The Monotonicity Principle for nonlinear materials}
The Monotonicity Principle Method is an imaging method widely used to solve the inverse obstacle problem across several different fields. It was originally introduced for static problems (such as ERT, ECT, and MIT) \cite{Tamburrino_2002, Tamburrino2003233, harrach2015resolution, harrach2013monotonicity, harrach2018monotonicity}, and later employed for quasi-static problems, such as eddy currents in the low- and high-frequency limits, and time-domain eddy currents \cite{Tamburrino_2006,Su_2017, Su2017, Arnold_2013, ventre2016design, Tamburrino_2010, Tamburrino2015159, tamburrino2021themonotonicity,Tamburrino20161,Tamburrino201226,Tamburrino2006FastMF}. Furthermore, it has been successfully applied to wave propagation phenomena governed by the Helmholtz equation in both bounded and unbounded domains \cite{griesmaier2018monotonicity, harrach2019monotonicity, harrach2019dimension, albicker2020monotonicity, albicker2023monotonicity, AT_WAVE2015}, as well as to problems in linear elasticity \cite{eberle2020shape,eberle2020lipschitz}, and fractional diffusion \cite{harrach2019monotonicity-based, harrach2020monotonicity-based}. 

These methods rely on a monotonicity principle inherent to the underlying physics, which establishes a monotonic relationship between the spatial distribution of the unknown material property within the tomographic domain and the resulting boundary measurements. Recently, the Monotonicity Principle has been generalized to materials characterized by a nonlinear constitutive relationship. In particular, an MP was established in \cite{corboesposito2021monotonicity,corboesposito2024piecewise} for general nonlinear electrical conductivities in the form
\begin{equation}\label{int_mp_1}
    \sigma_1 \leq \sigma_2 \Longrightarrow \overline{\Lambda}_{\sigma_1}\leqslant\overline{\Lambda}_{\sigma_2}, 
\end{equation}
where the inequality $\sigma_1 \leq \sigma_2$ means that $\sigma_1(x,E)\leq\sigma_2(x,E)$ holds for a.e. $x\in\Omega$ and for all $E>0$, while $\overline{\Lambda}_{\sigma_1}\leqslant \overline{\Lambda}_{\sigma_1}$ stands for $\langle \overline{\Lambda}_{\sigma_1}(f),f\rangle\leq \langle \overline{\Lambda}_{\sigma_2}(f),f\rangle$, $\forall f$. Here $\overline{\Lambda}_{\sigma}$ denotes a specific nonlinear boundary operator, called the average DtN (A-DtN) operator, which plays a key role in this framework: as shown in \cite{corboesposito2021monotonicity}, it is the proper operator to establish an MP in the presence of general nonlinear electrical conductivities, unlike the \lq\lq classical\rq\rq \ DtN operator.

The MP in \eqref{int_mp_1} can easily be customized to the inverse obstacle problem. Indeed, from \eqref{int_mp_1} it follows that:
\begin{equation}
\label{eqn:mono2}
    T \subseteq A \Subset \Omega\Longrightarrow \overline{\Lambda}^{T}\leqslant\overline{\Lambda}^{A},
\end{equation}
or, equivalently,
\begin{equation}
\label{eqn:mono3}
\overline{\Lambda}^{T}	\nleqslant \overline{\Lambda}^{A}\Longrightarrow T\nsubseteq A
\end{equation}
where $A$ is the unknown anomalous region and $T$ an otherwise arbitrary set termed test set, $\overline{\Lambda}_A$ and $\overline{\Lambda}_T$ are the corresponding A-DtN operators, and the material property $\gamma$ in $A$ is assumed to be higher than that in the background $\Omega \setminus A$.

As first recognized in \cite{Tamburrino_2002} for linear constitutive relationships, \eqref{eqn:mono3} is the key to the imaging method, since it allows one to infer, starting only from boundary data $\overline{\Lambda}_A$ and $\overline{\Lambda}_T$, whether $T$ is not included in the anomalous region $A$. Building upon these theoretical advancements, the first MP-based imaging method specifically tailored for nonlinear inclusions was introduced in \cite{mottola2024imaging}.

Imaging methods based on \eqref{eqn:mono3} have been shown to possess unique features when compared to other imaging approaches. Indeed, MP is capable of perfectly reconstructing the boundary of the obstacle for noiseless data (see \cite{harrach2013monotonicity} for the linear case and \cite{mottola2025theinverseobstacle,mottola2026corrigendumtheinverseobstacle} for the nonlinear one), and the reconstruction is intrinsically stable with respect to noise, provided that an appropriate regularization strategy is applied (see \cite{garde2017convergence} for the linear case and \cite{mottola2025theinverseobstacle,mottola2026corrigendumtheinverseobstacle} for the nonlinear one).

Currently, although the MP proven in \cite{corboesposito2021monotonicity, corboesposito2024piecewise} holds for general nonlinear constitutive relationships for both the background and the anomalous phase, its practical implementation in \cite{mottola2024imaging} applies only to linear backgrounds, while allowing for general constitutive relationships within the anomaly.

\subsection{$p-$Laplace and quasi $p-$Laplace materials}
The study of MP in the presence of nonlinear materials was initially conducted for the $p-$Laplacian variant of the Calderon problem. The foundational formulation of the inverse problem dates back to \cite{Salo2012_IP}, where the boundary value of the $p-$Laplace conductivity was shown to be uniquely determined by a nonlinear DtN map. The uniqueness was subsequently extended to the first-order derivative \cite{brander2016calderon}. 

Regarding imaging methods, an inversion algorithm leveraging the enclosure method for the $p$-Laplacian was developed in \cite{brander2015enclosure} to recover the convex hull of an inclusion. 
Concurrently, a tailored version of the MP for the $p$-Laplacian framework was derived in \cite{brander2018monotonicity, guo2016inverse}, with \cite{brander2018monotonicity} proving that the theoretical reconstruction limit corresponds to the convex hull of the anomalies. Both the Enclosure and the MP method can be applied only when the constitutive relationships for the background and the anomaly are of monomial type and share the same exponent. Extensions to include the linear terms have also been proposed (quasi $p-$Laplace variants).

Moreover, \cite{carstea2020recovery} established a uniqueness proof and a reconstruction procedure for the specific Calderón problem where the nonlinearity is the sum of a linear and a $p$-Laplacian term.

\subsection{The $p-$Laplace Signature}
\label{pLS_sec}
A crucial role in this paper is played by the \emph{p-Laplace Signature} (pLS), recently studied in \cite{corboesposito2024theplaplacesignature,corboesposito2024thep0laplacesignature}.

\begin{figure}[ht]
    \centering
\includegraphics[width=0.7\textwidth]{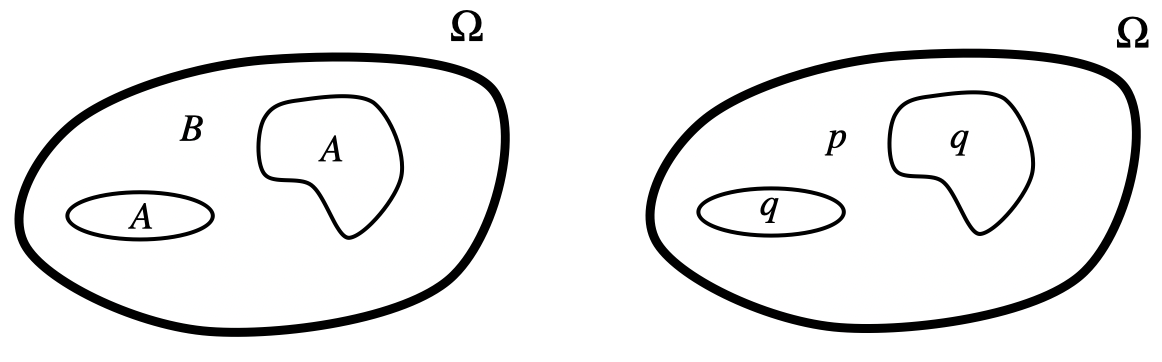}
    \caption{A two-phase configuration (left) where $p$ and $q$ are the growth exponents at large $s$ for regions $A$ and $B$, respectively (right). The picture is adapted from \cite[Figure 3]{corboesposito2024thep0laplacesignature} and it is courtesy of Siam Journal of Imaging Science.}
    \label{fig_02_large}
\end{figure}

Let $\gamma_a$ and $\gamma_b$ be the material properties in regions $A \Subset \Omega$ and $B = \Omega \setminus A$, respectively, and let the material properties satisfy the following asymptotic behavior (see Figure \ref{fig_02_large}):
\begin{align}
\label{eq:asy1}
\gamma_b(x,s) \sim \beta (x)s^{p-2} \quad
\text{for } s \to +\infty,\\
\label{eq:asy2}
\gamma_a(x,s) \sim \alpha (x)s^{q-2} \quad \text{for }s \to +\infty.
\end{align} 

The key result proved in \cite{corboesposito2024theplaplacesignature} is that when the boundary data is of the type $\lambda f$, where $\lambda$ is a scalar and $f$ an arbitrary Dirichlet data, it result that the rescaled solution $v^\lambda = u^\lambda / \lambda$, where $u^\lambda$ solves \eqref{problem} with boundary data $\lambda f$, approaches a limit $v^\infty$, i.e.
\begin{equation}
    \lim_{\lambda \to +\infty} \frac{u^{\lambda}}{\lambda} = v^\infty.
\end{equation}
For example, in the specific case of steady-state currents, $u^{\lambda}$ represents the scalar potential, $\mathbf{E}^{\lambda} = - \nabla u^{\lambda}$ is the electric field, and $\mathbf{J}^{\lambda}(x)=\gamma(x, \left|  \mathbf{E}^{\lambda} \right|) \mathbf{E}^{\lambda}$ is the electric current density, for $\lambda \to +\infty$ it follows that
\begin{align}
    u^{\lambda}(x) & \sim \lambda v^\infty(x), \\
    \mathbf{E}^{\lambda} & \sim \lambda \mathbf{e}^{\infty}, \\
    \mathbf{J}_b^{\lambda} & \sim \lambda^{p-1} \mathbf{j}^{\infty}_b, \\
    \mathbf{J}_a^{\lambda} & \sim \lambda^{q-1} \mathbf{j}^{\infty}_a,
\end{align}
where
\begin{align}
    \mathbf{e}^{\infty} & = - \nabla v^\infty, \\
    \mathbf{j}_b^{\infty} & = \beta (x)  \left| \mathbf{e}^{\infty} \right|^{p-2}  \mathbf{e}^{\infty}, \\
    \mathbf{j}_a^{\infty} & = \alpha (x)  \left| \mathbf{e}^{\infty} \right|^{q-2}  \mathbf{e}^{\infty}.
\end{align}
The normalized solution $v^\infty$ solves a proper {weighted} $p-$Laplace problem in region $B$ and a weighted $q-$Laplace problem in region $A$, with proper transmission conditions between $A$ and $B$.

From the physical standpoint, for $\lambda$ large enough, region $A$ acts as a perfect electric conductor for $p<q$ and as a perfect electric insulator for $p>q$:
\begin{itemize}
    \item if $p<q$, region $A$ acts as a \emph{perfect electric conductor} (PEC);
    \item if $p>q$, region $A$ acts as a \emph{perfect electric insulator} (PEI).
\end{itemize}

The case $p = q$ corresponds to $A = \emptyset$.

The relevance of these results is particularly evident in the case $p=2$. In this setting, regardless of the nonlinear constitutive law of the original materials, the large-data limit is always described by a linear problem.

\begin{figure}[ht]
    \centering
\includegraphics[width=0.7\textwidth]{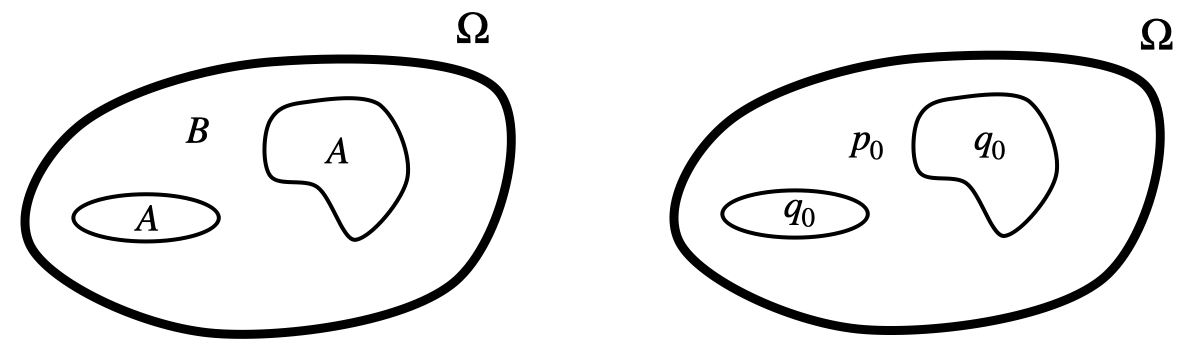}
    \caption{A two-phase configuration (left) where $p_0$ and $q_0$ are the growth exponents at small $s$ for regions $A$ and $B$, respectively (right). The picture is adapted from \cite[Figure 3]{corboesposito2024thep0laplacesignature} and it is courtesy of Siam Journal of Imaging Science.}
    \label{fig_03_small}
\end{figure}

A similar analysis is also available for the small data regime ($\lambda \to 0^+$) \cite{corboesposito2024thep0laplacesignature}. In this case, the material properties $\gamma_a$ and $\gamma_b$ satisfy the following asymptotic behavior for $s \to 0^+$ (see Figure \ref{fig_03_small}):
\begin{align}
\label{eq:asys}
\gamma_b(x,s) \sim \beta_0 (x)s^{p_0-2} \quad
\text{for } s \to 0^ +,\\
\label{eq:asysm}
\gamma_a(x,s) \sim \alpha_0 (x)s^{q_0-2} \quad \text{for }s \to 0^+.
\end{align} 

\subsection{New classes of nonlinearity tractable via the A-MPM}

The main goal of this work is to extend the A-MPM---the most advanced tool currently available---to inverse obstacle problems involving nonlinear materials. 

Evaluating the theoretical reconstruction limits in a nonlinear background is highly challenging. To frame this problem, four key concepts are essential:
\begin{itemize}
    \item \textbf{Well-Separated (WS):} the material properties for the anomaly $\gamma_a$ and the background $\gamma_b$ are well-separated if and only if either $\gamma_a(x,s)>\gamma_b(x,s)$ or $\gamma_b(x,s)>\gamma_a(x,s)$, for a.e. $x \in \Omega$ and $\forall s>0$. 
    \item \textbf{Asymptotic Well-Separated (AWS):} the material properties for the anomaly $\gamma_a$ and the background $\gamma_b$ are well-separated if and only if either $\gamma_a(x,s)>\gamma_b(x,s)$ or $\gamma_b(x,s)>\gamma_a(x,s)$, for a.e. $x \in \Omega$ and for $s \to +\infty$ (or  $s \to 0^+$). 
    \item \textbf{Outer Support Method (OSM):} an imaging method is an outer boundary method if it is capable of reconstructing the outer support $A^*$ of the anomaly $A$.
    \item \textbf{Convex Hull Method (CHM):} an imaging method is a convex hull method if it is capable of reconstructing the convex hull of the anomaly $A$.
\end{itemize}
Currently, the A-MPM acts as an OSM for $p=2$ and arbitrary $q$ \cite{mottola2025theinverseobstacle,mottola2026corrigendumtheinverseobstacle}, whereas P-MPM and P-EC act as CHMs strictly when both the background and the anomaly are of $p$-Laplace type \cite{brander2018monotonicity,brander2015enclosure}. 

In this contribution, these boundaries are generalized. First, it is proved that the A-MPM guarantees the reconstruction of the anomaly's convex hull for a $p$-Laplace background when $q=p$, even if the anomaly itself is not $q$-Laplace. 

Second, it is demonstrated that the $p$-Laplace Signature (pLS) acts as a crucial bridge to imaging methods. By relying on extreme boundary data, the pLS provides an equivalent WS regime when only AWS holds. Combining the A-MPM with the pLS enables the reconstruction of the outer support of $A$ when the background is bounded ($p=2$) and $q=p$, or its convex hull when $p \neq 2$ and $q=p$. In all these scenarios, the convex hull represents the minimum guaranteed reconstruction, though better results may be achievable.

The new reconstruction strategy strictly depends on the material properties:
\begin{itemize}
    \item \textbf{Neither WS nor AWS:} Currently available methods fail to yield a reconstruction.
    \item \textbf{Only AWS:} The pLS is mandatory to replace the original problem with an equivalent WS formulation in the limit of extreme boundary data (see \Cref{sec_p=2_quasi} for $p=2$ and \Cref{pnot2_quasi} for $p \neq 2$).
    \item \textbf{Only WS:} The Monotonicity Principle can be applied directly, both for linear ($p=2$) and $p$-Laplace ($p\neq 2$) backgrounds (see \Cref{pnot2_Lap}).
    \item \textbf{Both WS and AWS:} Both approaches are valid, allowing the pLS to extend existing methods to a broader class of backgrounds.
\end{itemize}
All the results, previous and proposed in this contribution, are summarized in \Cref{tab_sum_1}.
The only scenario left uncovered by the present theory is when $p\neq 2$ and $q\neq p$, a limitation stemming from the lack of a converse monotonicity theory for the $p$-Laplace equation in the presence of PEC or PEI inclusions.
\begin{table}[htbp]
    \centering
    \begin{tabular}{|c|c|c|>{\raggedright\arraybackslash}m{9cm}|}
        \hline
            & \textbf{WS} & \textbf{AWS} & \multicolumn{1}{c|}{\textbf{Theoretical results}} \\
        \hline\hline
        \#1 & $\circ$ & $\circ$ & 
\begin{tabular}{@{}r@{\;}>{\raggedright\arraybackslash}p{8.2cm}@{}}
& \quad\, \textbf{N/A} \end{tabular}\\
        \hline
        \#2 & $\circ$ & $\bullet$ & 
        \begin{tabular}{@{}r@{\;}>{\raggedright\arraybackslash}p{8.2cm}@{}}
            (a) & \textbf{A-MPM {by pLS}} ($p=2$ quasilinear background, $\forall \, q$ quasilinear anomaly; OSM) \\
            (b) & \textbf{A-MPM {by pLS}} ($p\neq 2$ quasilinear background; $q=p$ quasilinear anomaly; CHM); \\
        \end{tabular} \\ 
        \hline
        \#3 & $\bullet$ & $\circ$ & 
        \begin{tabular}{@{}r@{\;}>{\raggedright\arraybackslash}p{8.2cm}@{}}
            (a) & \textbf{A-MPM} ($p=2$ linear background, $\forall \, q$ quasilinear anomaly; OSM \cite{mottola2025theinverseobstacle,mottola2026corrigendumtheinverseobstacle}) \\
            (b) & \textbf{A-MPM} ($p\neq 2$ $p-$Laplacian background; $p=q$ quasilinear anomaly; CHM);
        \end{tabular} \\
        \hline
        \#4 & $\bullet$ & $\bullet$ & 
        \begin{tabular}{@{}r@{\;}>{\raggedright\arraybackslash}p{8.2cm}@{}}
            (a) & \textbf{A-MPM {by pLS}} ($p=2$ quasilinear background, $\forall \, q$ quasilinear anomaly; OSM) \\
            (b) & \textbf{A-MPM {by pLS}} ($p\neq 2$ quasilinear background; $q=p$ quasilinear anomaly; CHM); \\
            (c) & \textbf{A-MPM} ($p=2$ linear background, $\forall \, q$ quasilinear anomaly; OSM \cite{mottola2025theinverseobstacle,mottola2026corrigendumtheinverseobstacle}) \\
            (d) & \textbf{A-MPM} ($p\neq 2$ $p-$Laplacian background; $p=q$ quasilinear anomaly; CHM); \\
            (e) & \textbf{P-EM} ($p=q$, $p-$Laplacian background and anomaly, CHM, \cite{brander2015enclosure}); \\
            (f) & \textbf{P-MPM} ($p=q$, $p-$Laplacian background and anomaly, CHM, \cite{brander2018monotonicity});  \\
        \end{tabular} \\
        \hline
    \end{tabular}
    \caption{Overview of the available theoretical results based on the properties of the nonlinear phases. Cases without a reference are the new one proposed in this work.}
    \label{tab_sum_1}
\end{table}



\section{Framework of the Problem}
\label{notation_sec}
\subsection{Notations}

Throughout this paper, $\Omega$ denotes the region occupied by the conducting materials. It is assumed that $\Omega\subset\R^n$, $n\geq 2$, is a bounded domain (i.e. an open and connected set) with Lipschitz continuous boundary and $A\Subset\Omega$ is an open bounded set with Lipschitz boundary and a finite number of connected components, such that $B:=\Omega\setminus\overline A$ is still a domain. Hereafter, it is assumed that the anomaly occupies the set $A\Subset\Omega$, while the background material occupies region $B$ (see Figure \ref{fig_01_BmenoA}). Summarising, $A \in \mathcal{S}(\Omega)$, where
\begin{equation}\label{SOmega}
\begin{split}
\mathcal{S} (\Omega):= & \left\{ V \subseteq \Omega: V \text{ is an open bounded set with a Lipschitz boundary} \right. \\ 
&\  \left. \text{and } \partial V \text{ is made by a finite number of connected components}  \right\}.
\end{split}
\end{equation}

Moreover $dx$ and $dS$ denote the $n-$dimensional and the $(n-1)-$dimensional Hausdorff measure, respectively. Furthermore,
\[
L^\infty_+(\Omega):=\{\theta\in L^\infty(\Omega)\ |\ \theta\geq c_0\ \text{a.e. in}\ \Omega, \ \text{for a positive constant}\ c_0\}.
\]
Finally, for any $1<s<+\infty$, $W^{1,s}_0(\Omega)$ denotes the closure set of $C_0^1(\Omega)$ with respect to the $W^{1,s}-$norm.

The applied boundary voltage $f$ belongs to the abstract trace space $B_{p}^{1-\frac 1p,p}(\partial\Omega)$, which, for any bounded Lipschitz open set, is a Besov space (refer to \cite{JERISON1995161,leoni17}, and \cite[Section 2.1]{corboesposito2024piecewise}).

For the sake of brevity,  this space is denoted by $X^p(\partial \Omega)$ and its elements can be identified as the functions in $W^{1,p}(\Omega)$, modulo the equivalence relation $f\in [g]_{X^p(\partial \Omega)}$ if and only if $f-g\in W^{1,p}_0(\Omega)$, see \cite[Th. 18.7]{leoni17}.

Finally,  by $X^p_\diamond (\partial \Omega)$ is denoted the set of elements in $X^p(\partial \Omega)$ with zero average on $\partial\Omega$ with respect to the measure $dS$ and by $X^p_\diamond(\partial\Omega)'$ the dual space of $X^p_\diamond (\partial \Omega)$. 

\subsection{The assumptions} \label{sub_sec_assumptions}
Throughout the paper, a general nonlinear constitutive relationship of the form
 \begin{equation} \label{gOhm}
 {\bf T} (x, {\bf s})=\gamma (x, s) \bf s,
 \end{equation}
is considered for the isotropic material properties in the anomalous and background regions. In \eqref{gOhm},
(i) ${\bf s}={\bf E}$ is the electric field, ${\bf T}={\bf J}$ is the electrical current density, and $\gamma=\sigma$ is the electrical conductivity in the case of ERT; (ii) ${\bf s}={\bf E}$ is the electric field, ${\bf T}={\bf D}$ is the electric flux density, and $\gamma=\varepsilon$ is the electrical permittivity in the case of ECT; and (iii) ${\bf s}={\bf H}$ is the magnetic field, ${\bf T}={\bf B}$ is the magnetic flux density, and $\gamma=\mu$ is the magnetic permeability in the case of MIT (see also \cite{mottola2024imaging} for further details).

Hereafter, the assumptions required for the material property $\gamma$ are listed. Such assumptions are introduced to guarantee the well-posedness of the problem \eqref{problem}, which in turn is the minimal requirement to formulate the associated inverse problem (see \cite{corboesposito2021monotonicity,corboesposito2024piecewise,mottola2026corrigendumtheinverseobstacle}). 

In the following, $\gamma_b: \Omega\times[0,+\infty)\to\R$ denotes the restriction of the material property to the background region, while $\gamma_a: \Omega\times[0,+\infty)\to\R$ denotes the restriction of the material property to the anomalous region (see also \Cref{fig_02_large}).

Firstly, the definition of the Carath\'eodory function is recalled.
\begin{definition}
$\gamma:\Omega\times[0,+\infty)\to\R$ is a Carath\'eodory function iff:
\begin{itemize}
\item $x\in\overline\Omega\mapsto \gamma(x,s)$ is Lebesgue-measurable for every $s\in[0,+\infty)$,
\item $s\in [0,+\infty)\mapsto \gamma(x, s)$ is continuous for almost every $x\in\Omega$.
\end{itemize}
\end{definition}

\begin{itemize}
\item[{\bf (A)}] $\gamma_b$ and $\gamma_a$ are Carath\'eodory function. 
\item[{\bf (P1)}] For fixed $1<p<+\infty$, there exist two positive constants $\overline{\gamma}$ and $s_0$ 
such that: 
\[
\begin{split}
&\gamma_{b}(x,s)\leq \overline{\gamma}\left[1+\left( \frac{s}{s_0} \right)^{p-2}\right]\qquad \text{if}\ p\ge 2,\\
&\gamma_{b}(x,s)\leq \overline{\gamma}\left( \frac{s}{s_0} \right)^{p-2}\qquad\qquad\quad\   \text{if}\ 1<p< 2,\\
\end{split}
\]
$\text{for a.e.}\ x\in {\overline B}\ \text{and}\ \forall s\ge 0$.
\item[{\bf (P2)}] For fixed $1<p <+\infty$, there exist two positive constants $\underline\gamma$ and $s_0$ such that:
\begin{equation*}
\begin{split}
\biggl(\gamma_b(x,s_2)&\frac{{\bf s}_2}{s_0}-\gamma_b(x,s_1)\frac{{\bf s}_1}{s_0}\biggl)\cdot\left( \frac{{\bf s}_2}{s_0}-\frac{{\bf s}_1}{s_0}\right)\\
        &\geq
        \begin{cases}\displaystyle \underline\gamma\left|\frac{{\bf s}_2}{s_0}-\frac{{\bf s}_1}{s_0}\right|^p\ &\text{if} \ p\geq 2\\ \displaystyle
        \underline\gamma\left(1+ \left|\frac{{\bf s}_2}{s_0}\right|^2+\left|\frac{{\bf s}_1}{s_0}\right|^2\right)^\frac{p-2}2\left|\frac{{\bf s}_2}{s_0}-\frac{{\bf s}_1}{s_0}\right|^2\ &\text{if}\ 1<p<2
        \end{cases}
        \\
        \end{split}
       \end{equation*}
       \ $\text{for a.e.}\ x\in B$, and for any ${\bf s}_1,{\bf s}_2\in\R^n$.
\end{itemize}

\begin{itemize}
\item[{\bf (Q1)}] For fixed $1<q<+\infty$, there exist two positive constants $\overline{\gamma}$ and $s_0$
such that:  
\[
\begin{split}
&\gamma_{a}(x,s)\leq \overline{\gamma}\left[1+\left( \frac{s}{s_0} \right)^{q-2}\right] \qquad \text{if}\ q\ge 2,\\ &\gamma_{a}(x,s)\leq \overline{\gamma}\left( \frac{s}{s_0} \right)^{q-2}\qquad\qquad\quad\ \text{if}\ 1<q< 2,
\end{split}
\]
$\text{for a.e.}\ x\in A\ \text{and}\ \forall s>0$.
\item[{\bf (Q2)}] For fixed $1<q<+\infty$, there exist two positive constants $\underline{\gamma}$ and $s_0$ such that: 
\begin{equation*}
\begin{split}\biggl(\gamma_a(x,s_2)&\frac{{\bf s}_2}{s_0}-\gamma_a(x,s_1)\frac{{\bf s}_1}{s_0}\biggl)\cdot\left(\frac{{\bf s}_2}{s_0}-\frac{{\bf s}_1}{s_0}\right)      \\ 
&\geq
        \begin{cases}
\displaystyle\underline{\gamma}\left|\frac{{\bf s}_2}{s_0}-\frac{{\bf s}_1}{s_0}\right|^q\ &\text{if} \ q\geq 2\\
            \displaystyle\underline\gamma\left(1+ \left|\frac{{\bf s}_2}{s_0}\right|^2+\left|\frac{{\bf s}_1}{s_0}\right|^2\right)^\frac{q-2}2\left|\frac{{\bf s}_2}{s_0}-\frac{{\bf s}_1}{s_0}\right|^2\ &\text{if}\ 1<q<2
        \end{cases}
        \\
\end{split}
\end{equation*}$\ \text{for a.e.}\ x\in A,$
for any ${\bf s}_1,{\bf s}_2\in\R^n$.

\item[{\bf (R)}] $\gamma_{a}(x, s)=0$,  $\text{for a.e.}\ x\in A$.
\item[{\bf (S)}] $\gamma_{a}(x,s)=+\infty$,  $\text{for a.e.}\ x\in A$.
\end{itemize}

Assumptions (QX), (R) and (S) are alternative. In the case $A=\emptyset$, the assumptions (QX), (R) and (S) are not considered.

\subsection{Foundation of the problem}
The governing equation is given by \eqref{problem}, where $f\in X_\diamond^p(\partial \Omega)$. Problem \eqref{problem} is meant in the weak sense, that is
\begin{equation*}
\int_{\Omega }\gamma \left( x,| \nabla u(x) |\right) \nabla u(x) \cdot\nabla \varphi (x)\ \text{d}x=0\quad\forall\varphi\in C_c^\infty(\Omega).
\end{equation*}

The solution $u$ restricted to $B$ belongs to $W^{1,p}(B)$, whereas $u$ restricted to $A$ belongs to $W^{1,q}(A)$;  {however} the solution $u$ as a whole is an element of the  {largest between the two functional spaces $W^{1,p}(\Omega)$ and $W^{1,q}(\Omega)$. Furthermore, }(i) if $p\leq q$ then $W^{1,p}(\Omega)\cup W^{1,q}(\Omega)=W^{1,p}(\Omega)$, and (ii) if $q\leq p$ then $W^{1,p}(\Omega)\cup W^{1,q}(\Omega)=W^{1,q}(\Omega)$.

The solution $u$ satisfies the boundary condition in the sense that $u-f\in W_0^{1,p}(\Omega)\cup W_0^{1,q}(\Omega)$ and it is written $u|_{\partial\Omega}=f$.

For the two-phase material problem considered in this work, Equation \eqref{problem} can be cast as
\begin{equation}\label{P1}
\begin{cases}
\dive\left(\gamma_{b}\left(x,|\nabla u(x)|\right)\nabla u(x)\right)=0 & \text{in $\Omega\setminus A$} \\
\dive\left(\gamma_{a}\left(x,|\nabla u(x)|\right)\nabla u(x)\right)=0 & \text{in $A$} \\
u(x)=f(x) & \text{on $\partial\Omega$} \\
\gamma_{b}\left(x^+,|\nabla u(x^+)|\right)\partial_n v(x^+)=\gamma_{a}\left(x^-,|\nabla u(x^-)|\right)|\partial_nu(x^-) & \text{on $\partial A$} \\
u(x^+)=u(x^-) & \text{on $\partial A$},
\end{cases}
\end{equation}
where $u(x^+)$, $u(x^-)$, $\partial_n u(x^+)$ and $\partial_n u(x^-)$ are meant as the limits in $x$ evaluated from the outer $(+)$ or from the inner $(-)$ side of the boundary of $A$.

Moreover, a minimizer $u$ of the variational problem
\begin{equation}
\label{gminimum}
\min\left\{ \mathbb{E}^A\left( \xi\right)\ :\ \xi\in W^{1,p}(\Omega)\cup W^{1,q}(\Omega), \ \xi|_{\partial\Omega}=f\right\},
\end{equation}
solves \eqref{P1}.
In (\ref{gminimum}), the functional $\mathbb{E}^A\left( \cdot\right)$ is the Dirichlet energy
\begin{equation}
\label{genergy}
\mathbb{E}^A
\left(  \xi \right) =\int_{\Omega} Q^A (x,|\nabla \xi(x)|)\ \text{d}x= \int_{B} Q_b (x,|\nabla \xi(x)|)\ \text{d}x+ \int_A Q_a (x,|\nabla \xi(x)|)\ \text{d}x
\end{equation} 
where $Q_b$ and $Q_a$ are the Dirichlet energy density in $B$ and in $A$, respectively:
\begin{align*}
& Q_{b} \left( x,s\right)  :=\int_{0}^{s} \gamma_b\left( x,\eta \right)\eta\text{d}\eta\quad \text{for a.e.}\ x\in B\ \text{and}\ \forall s\geq0,\\
& Q_{a}\left( x,s\right)  :=\int_{0}^{s} \gamma_a\left( x,\eta \right)\eta  \text{d}\eta\quad \text{for a.e.}\ x\in A\ \text{and}\ \forall s\geq 0.
\end{align*}

Furthermore, if $\gamma_a(x,s)=0$, the corresponding problem is
\begin{equation}
\label{problem_PEI}
\begin{cases}
\dive (\gamma_b(x, |\nabla u|)\nabla u)=0 & \text{in } B \\
\gamma_b(x, |\nabla u|)\partial_\nu u=0 & \text{on } \partial A\\
u\in W^{1,p}(B), \ 
u=f & \text{on }\partial\Omega
\end{cases}
\end{equation}
which in the case of ERT, i.e. $\gamma=\sigma$, corresponds to have a PEI (Perfect Electric Insulator) in region $A$.
The solution $u$ is variationally defined as the minimum of the following Dirichlet Energy
\begin{equation}
\label{minimum_PEI}
\min_{\substack{\xi\in W^{1,p}(B)\\ \xi=f \text{ on }\partial\Omega}}\mathbb E^A (\xi),
\end{equation}
where
\begin{equation}
    \label{genergy_outer}
\mathbb{E}^A(\xi)=\int_{B} Q_{b}(x,|\nabla \xi(x)|)dx.
\end{equation}
It is worth noticing that, since $\gamma_a(x,s)=0$, the integration domain can be extended to the whole domain $\Omega$, so that \eqref{genergy_outer} can be viewed as a particular case of the general energy \eqref{genergy}.

Similarly, if $\gamma_a(x,s)=+\infty$, the corresponding problem is
\begin{equation}
\label{problem_PEC}
\begin{cases}
\dive (\gamma_b(x, |\nabla u|)\nabla u)=0 & \text{in }B \\
|\nabla u|=0 & \text{a.e. in }A\\
\int_{\partial A_i}\gamma_b(x,|\nabla u(x)|)\partial_\nu u(x)dS=0 & i=1,...,M\\
u\in W^{1,p}(\Omega), \ u=f & \text{on }\partial\Omega
\end{cases}
\end{equation}
which corresponds to a perfectly conducting conductor (PEC) occupying region $A$, in the case of ERT. 
The solution $u$ is variationally defined as the minimum of the following Dirichlet Energy
\begin{equation}
\label{minimum_PEC}\min_{\substack{\xi\in W^{1,p}(\Omega)\\ \xi=f \text{ on } \partial\Omega\\ |\nabla \xi|=0\text{ in } A}}\mathbb E^A(\xi),
\end{equation}
where $\mathbb{E}^A$ is again given by \eqref{genergy_outer}. Let us stress that, since the solution in \eqref{minimum_PEC} assumes a constant value in $A$, then again the integration domain can be extended to the whole domain $\Omega$.

\subsection{The boundary operators}\label{sec_bou}
\label{sec:dtns}
The Dirichlet-to-Neumann (DtN) operator maps the Dirichlet data into the corresponding Neumann data:
\begin{equation*}
\Lambda^A  :f\in X_\diamond^p(\partial\Omega)\mapsto \gamma^A(x, |\nabla u|)\ 
\partial_nu|_{\partial\Omega} 
\in X_\diamond^p(\partial\Omega)',
\end{equation*}
where 
\begin{equation}
\label{gammaA}
\gamma^A(x,s)=\begin{cases}
   \gamma_b(x,s) & \text {in $\Omega\setminus A$} \\
   \gamma_{a}(x,s) & \text {in $A$}.
  \end{cases}
\end{equation}
for any $s \geq 0$.

From a physical point of view, the DtN operator maps the imposed boundary data to the quantity measured on the boundary $\partial\Omega$. For instance, in ERT the DtN maps the imposed boundary electric scalar potential to the normal component of the electrical current density entering $\partial \Omega$ (see \cite{mottola2024imaging} for a detailed discussion in the case of MIT and ECT).

In weak form, the DtN operator is
\begin{equation}
\label{w-DtN}
\langle \Lambda^A  \left( f\right) ,\psi\rangle
=\int_{\partial \Omega }\psi (x) \gamma^A\left( x, \left\vert \nabla u(x)\right\vert\right)  \partial_n u(x)\,\text{d}S\quad\forall \psi\in X^p_\diamond(\partial\Omega).
\end{equation}

Furthermore, by testing the DtN operator \eqref{w-DtN} with the solution $u$ of \eqref{P1} and using a divergence Theorem, it results 
\begin{equation}\label{w-DtN-f}
\langle \Lambda^A  \left( f\right) ,f\rangle
=\int_{\Omega } \gamma^A( x ,\nabla u(x)){ |\nabla u}(x)|^2\ \text{d}x.
\end{equation}

The Average DtN (A-DtN) is defined as (see \cite{corboesposito2024piecewise,corboesposito2021monotonicity})
\begin{equation}
\label{P4}
\overline{\Lambda}^A: f\in X_{\diamond} ^p(\partial\Omega) \to \int_0^1\Lambda^A(\alpha f)\,d\alpha\in X_\diamond^p(\partial\Omega)',
\end{equation}
where
\begin{equation}
\label{P3}
    \Lambda^A:f\in X_{\diamond}^p(\partial\Omega)\to \gamma^A\partial_n u^A|_{\partial\Omega}\in X_\diamond^p(\partial\Omega)',
\end{equation}
is the classical DtN operators related to anomalies occupying regions $A$.

\begin{notation}
    Hereafter, $\gamma^V$ denotes a material property of the form \eqref{gammaA}, for an anomaly occupying region $V$, i.e. $\gamma^V$ restricted to $V$ is given by $\gamma_a$. Coherently, $u^V$ is the solution of problem \eqref{P1} (equivalently the minimum of problem \eqref{gminimum}) for $\gamma=\gamma^V$, and $\Lambda^V$, $\overline{\Lambda}^V$ are the corresponding DtN and A-DtN operators, respectively.
\end{notation}

\section{The Monotonicity Principle and the $p$-Laplace Signature}\label{sec_monotonicity_principle}

In this section, the two main theoretical results underlying this work are recalled: the Monotonicity Principle (MP) \cite{corboesposito2021monotonicity,corboesposito2024piecewise} and the p-Laplacian Signature (pLS) \cite{corboesposito2024thep0laplacesignature, corboesposito2024theplaplacesignature}, concerning the asymptotic behaviours with \lq\lq large'' or \lq\lq small'' boundary data.

\subsection{The Monotonicity Principle}
\label{theMP}

The A-DtN operator introduced in \Cref{sec_bou} plays a key role, as it provides a direct link between a measurable boundary quantity and an internal quantity, namely the Dirichlet Energy. Specifically, the Dirichlet Energy can be transferred to a boundary measurement involving $\overline{\Lambda}^A$ as follows, see \cite[Th. 5.3]{corboesposito2024piecewise}.

\begin{theorem}\label{transferthm}
Let $1<p,q<+\infty$ and $\gamma^A$ satisfying {\bf (A)}, {\bf (P1)} and {\bf (P2)}, then
\begin{equation}
\label{ADtN=E}
\mathbb{E}^A\left( u^A\right)=\left\langle\overline{\Lambda}^A\left( f\right) ,f \right\rangle 
\quad \forall f\in X_\diamond^p(\partial\Omega),
\end{equation}
where 
\begin{itemize}
\item[(i)] if $A=\emptyset$, then $u^A$ is the minimizer of \eqref{gminimum};
\item[(ii)] if ${\gamma}^A$ satisfies {\bf (Q1)}-{\bf (Q2)}, then $u^A$ is the minimizer of \eqref{gminimum};
\item[(iii)] if ${\gamma}^A$ satisfies {\bf (R)}, then then $u^A$ is the minimizer of \eqref{minimum_PEI};
\item[(iv)] if ${\gamma}^A$ satisfies {\bf (S)}, then then $u^A$ is the minimizer of \eqref{minimum_PEC}.
\end{itemize}
\end{theorem}

Furthermore, the Monotonicity Principle for the average DtN $\overline{\Lambda}$ holds (see \cite[Th. 5.3]{corboesposito2024piecewise}).
\begin{theorem}
\label{monothm}
Let $1<p,q<+\infty$ and $\gamma^A$, $\gamma^T$ satisfying {\bf (A)}, {\bf (P1)} and {\bf (P2)}, then
\begin{equation}
\label{m_charge}
\gamma^T\leq\gamma^A\quad\Longrightarrow\quad \left\langle\overline{\Lambda}^T\left( f\right) ,f \right\rangle
\leq \left\langle\overline{\Lambda}^A \left( f\right) ,f \right\rangle
\quad \forall f\in X_\diamond^p(\partial\Omega),
\end{equation}
and one of the following holds
\begin{enumerate}
    \item[(i)] $A=\emptyset$;
    \item[(ii)] $\gamma^T,\gamma^A$ satisfy {\bf (Q1)}-{\bf (Q2)}; 
    \item[(iii)] $\gamma^T, \gamma^A$ satisfy {\bf (R)}; 
    \item [(iv)] $\gamma^T, \gamma^A$ satisfy {\bf (S)}.
\end{enumerate}
\end{theorem}
In \eqref{m_charge}, $\gamma^T\leq \gamma^A$  is meant in the sense that
\begin{equation}
\label{defmon}
\gamma^T(x,s)\leq \gamma^A(x,s) \quad \text{for a.e.}\ x\in\overline\Omega\ \text{and}\ \forall\ s>0.
\end{equation}

\subsubsection{The reconstruction method}\label{sec_met_mp}
It is worth noticing that for a two-phase material problem, the result of Theorem \ref{monothm} can be recast equivalently as in \eqref{eqn:mono2}, where $A$ is the targeted anomalous region, and $T$ is a test region to be determined if it is (not) included in region $A$.
The imaging strategy relies on the following equivalent form of \eqref{eqn:mono2}:
\begin{equation}\label{eqn:mono3r}
    \overline{\Lambda}^{T}	\nleqslant \overline{\Lambda}^{A}\Longrightarrow T\nsubseteq A.
\end{equation}

Relation \eqref{eqn:mono3r} provides a criterion to determine whether a test domain $T$ is not contained in the unknown anomaly $A$ using boundary measurements only. More precisely, the inequality $\overline{\Lambda}^{T}	\nleqslant \overline{\Lambda}^{A}$ provides a sufficient condition to conclude that $T \nsubseteq A$. Repeating the monotonicity test \eqref{eqn:mono3r} over a collection of test domains covering the region of interest yields a reconstruction method, where the estimate $A^{\dag}$ of the inclusion $A$ is
\begin{align}  \label{eqn:recon}
  A^{\dag}&=\bigcup \left\{T \in \mathcal S(\Omega)\,|\,\overline{\Lambda}^{A}-\overline{\Lambda}^{T}\geqslant 0\right\}. 
\end{align}

It is worth noting that $A^{\dag}$ is an upper bound to $A$, i.e. $A \subseteq A^{\dag}$, because $A \in \mathcal S(\Omega)$.

The reconstruction method of \eqref{eqn:recon} is valid in the absence of noise, but in any practical application, noise corrupts the data. As discussed in detail in \cite{garde2017convergence,garde2019regularized} for linear coefficients, and \cite{mottola2025theinverseobstacle,mottola2026corrigendumtheinverseobstacle} for quasilinear anomalies in a linear background, MP can be naturally regularized and stabilized to treat noisy data and modelling errors.

\subsection{The $p-$Laplace Signature}\label{sec_p_sig}
In this section, it is recalled from \cite{corboesposito2024theplaplacesignature, corboesposito2024thep0laplacesignature} that the nonlinear problem \eqref{genergy} can be approximated by a weighted $p-$Laplace problem. 
Specifically, for \lq\lq large\rq\rq\ or \lq\lq small\rq\rq\  Dirichlet data in the presence of two materials of different asymptotic growth exponents ($p\neq q$): 
\begin{itemize}
\item[(i)] the material filling the anomalous region can be replaced by either a perfect electric conductor or a perfect electric insulator;
\item[(ii)] the background material can be replaced by a material giving rise to a weighted $p-$Laplace problem.
\end{itemize}

\subsubsection{The \texorpdfstring{$p$-Laplace}{p-Laplace} Signature with large boundary data}
\label{large_sec}

This Section is devoted to the treatment of \lq\lq large'' boundary data, i.e. in the presence of an electric scalar potential applied to the boundary of the form $\lambda f$, where $\norm{f}=1$, and $\lambda\gg 1$.

In order to guarantee the convergence of the solution of problem \eqref{problem}, for \lq\lq large'' data, two additional assumptions are needed together with the ones specified in \Cref{sub_sec_assumptions}. Specifically,
\begin{itemize}
\item[{\bf (P$_\infty$3)}] For fixed $1<p<+\infty$, there exists a function $\beta$ such that:
\begin{equation*}
\lim_{s\to +\infty} \frac{\gamma_b (x,s)}{s^{p-2}}=\beta(x)\quad \text{for a.e.}\  x\in B.
\end{equation*}
\item[{\bf (Q$_\infty$3)}] For fixed $1<q<+\infty$, there exists a function $\alpha$ such that:
\begin{equation*}
\lim_{s\to +\infty} \frac{\gamma_a (x,s)}{s^{q-2}}=\alpha(x)\quad \text{for a.e.}\  x\in A.
\end{equation*}
\end{itemize}
where, $\beta\in L^\infty_+(B)$ and $\alpha\in L^\infty_+(A)$, by {\bf (P2)} and {\bf (Q2)}. A dedicated counterexample in \cite[Sec. 6]{corboesposito2024theplaplacesignature} show that the additional assumptions are sharp.

For the matter of convenience, a normalized solution $v^{A,\lambda}$ is introduced:
\begin{equation}
\label{reluv}
v^{A,\lambda}=\frac{u^{A,\lambda}}{\lambda},
\end{equation} 
where $u^{A,\lambda}$ is the minimizer of \eqref{gminimum} for a boundary data equal to $\lambda f$. Therefore, $v^{A,\lambda}$ is the solution of the following variational problem:
\begin{equation}
\label{G_norm}
\begin{split}
&\qquad\qquad\qquad\qquad\min_{\substack{\xi\in W^{1,p}(\Omega)\cup W^{1,q}(\Omega)\\ \xi=f\ \text{on}\ \partial \Omega}}\mathbb G^{A,\lambda}(\xi),\quad\\
&\mathbb G^{A,\lambda}(\xi)=\frac 1 {\lambda^p}\left(\int_{B} Q_b(x,\lambda|\nabla \xi(x)|)dx+\int_{A} Q_a(x,\lambda|\nabla \xi(x)|)dx\right).
\end{split}
\end{equation}
The multiplicative factor $1/\lambda^p$ is introduced in order to guarantee that the  minimum values of the functional $\mathbb G^{A,\lambda}$ are bounded for large $\lambda$.

Furthermore, the following relationship holds between the normalized and unnormalized Dirichlet Energy:
\begin{equation}
\label{relEG}
\mathbb E^A(\xi)=\lambda^p \mathbb G^{A,\lambda}\left(\frac { \xi}{\lambda}\right)\qquad\forall \xi\in W^{1,p}(\Omega)\cup W^{1,q}(\Omega).
\end{equation}

Since the nonlinear problem is approximated by a weighted $p-$Laplace model problem, the notation used to formulate the corresponding asymptotic result is introduced. Specifically, hereafter, $v^A$ denotes the solution of problem \eqref{problem_PEI} for $\gamma_b(x,s)=\beta(x)s^{p-2}$, i.e. for an anomalous region filled by a PEI and a monomial-type nonlinearity in the background material; equivalently (see also the beginning of the Section 5 in \cite{corboesposito2024theplaplacesignature}) $v^A$ is the minimizer of problem \eqref{minimum_PEI} when the energy $\mathbb E^A$ is replaced by
\begin{equation}\label{Benergy}
\mathbb B^A(\xi)=\int_{B} \beta(x)|\nabla \xi(x)|^pdx.
\end{equation}
Furthermore, $w$ denotes the solution of problem \eqref{problem_PEC} for $\gamma_b(x,s)=\beta(x)s^{p-2}$, i.e. for an anomalous region filled by a PEC and a monomial-type nonlinearity in the background material; equivalently $w$ is the minimizer of problem \eqref{minimum_PEC} when the energy $\mathbb E^A$ is replaced by \eqref{Benergy}.

The limiting behaviour of the normalized Dirichlet Energy for large boundary data is formalized in the following Theorem, but it is a byproduct of \cite[eq.s (5.14) and (5.25)]{corboesposito2024theplaplacesignature}. The case $p=q$ has not been treated explicitly in \cite{corboesposito2024theplaplacesignature} since it is easy to see that the Dirichlet Energy converges to a $p$-Laplace model energy with at most discontinuity in the coefficients.

\begin{theorem}
\label{Thm_conv_energy}
Let $1<q,p<+\infty$, $q\neq p$, $f\in X^p_\diamond(\partial \Omega)$, $\gamma^A$ satisfying {\bf (A)}, {\bf (P1)}, {\bf (P2)}, {\bf (P$_\infty$3)}, {\bf (Q1)}, {\bf (Q2)} and {\bf (Q$_\infty$3)}. Then
\begin{align}  
&\lim_{\lambda\to +\infty}\mathbb G^{A,\lambda}(v^{A,\lambda}) = \mathbb B^A(z^A)\quad \text{if } p>q,\\
&\lim_{\lambda\to +\infty}\mathbb G^{A,\lambda}(v^{A,\lambda}) = \mathbb B^A(w^A)\quad \text{if } p< q,\\
&\lim_{\lambda\to +\infty}\mathbb G^{A,\lambda}(v^{A,\lambda}) = \mathbb B^A(v^A)\quad \text{if } p= q,
\end{align}
where $v^{A,\lambda}\in W^{1,p}(\Omega)\cup W^{1,q}(\Omega)$ is the solution of \eqref{G_norm}, $z^A\in W^{1,p}(B)$
is the unique solution of
\eqref{minimum_PEI}; $w^A\in W^{1,p}(B)$ is the unique solution of \eqref{minimum_PEC}, for $\gamma_b(x,s)=\beta(x)s^{p-2}$; and $v^A\in W^{1,p}(\Omega)$ is the unique solution of \eqref{P1}, for $\gamma_b(x,s)=\beta(x)s^{p-2}$, and $\gamma_a(x,s)=\alpha(x)s^{p-2}$.
\end{theorem}
Theorem \ref{Thm_conv_energy} shows that, for $\lambda\to+\infty$, the Dirichlet Energy for the full quasilinear problem converges to the one of (i) a monomial background material property and a perfectly insulating inclusion if $p>q$, (ii) a monomial background material property and a perfectly conductive inclusion if $p<q$, and (iii) a $p$-Laplacian problem with a jump in the coefficients for $p=q$.

At this stage, building upon the previous convergence results for the energy, a new result is derived establishing the corresponding convergence of the A-DtN maps.

\begin{theorem}
\label{thm_conv_aDTN}
Let $1<p,q<+\infty$, $f\in X^{p}_\diamond(\partial \Omega)$, $\gamma^A$ and $\gamma^T$ satisfying {\bf (A)}, {\bf (P1)}, {\bf (P2)}, {\bf (Q1)}, {\bf (Q2)}, {\bf (P$_\infty$3)} and {\bf (Q$_\infty$3)}. Then
\begin{equation}
    \label{lim_ADtN_DtN_dim}
\lim_{\lambda\to +\infty}\frac{\left\langle\overline{\Lambda}^A\left( \lambda f\right) ,\lambda f \right\rangle}{\lambda^p}=\frac 1p \langle\Lambda_\infty^A(f),f\rangle,
\end{equation}
where
\begin{equation}
\label{lambda_3_cases}
\Lambda_\infty^A :f\in X^p_\diamond(\partial\Omega)\mapsto
\begin{cases}
\beta(x) \partial_n z^A|_{\partial\Omega} \in X^p_\diamond(\partial\Omega)'\quad &\text{if } p>q\\
 \beta(x) \partial_n w^A|_{\partial\Omega} \in X^p_\diamond(\partial\Omega)' &\text{if } p<q\\
\beta(x) \partial_n v^A|_{\partial\Omega} \in X_\diamond^p(\partial\Omega)'&\text{if } p=q\\
\end{cases},
\end{equation}
and $z^A\in W^{1,p}(B)$
is the unique solution of
\eqref{minimum_PEI}; $w^A\in W^{1,p}(B)$ is the unique solution of \eqref{minimum_PEC}, $v^A\in W^{1,p}(\Omega)$ is the unique solution of \eqref{P1}.
\end{theorem}
\begin{proof}
    By using \eqref{ADtN=E}, \eqref{relEG} and \eqref{reluv}, it holds:
\begin{equation}
\label{chainLGl_dim}
\left\langle\overline{\Lambda}^A\left( \lambda f\right) ,\lambda f \right\rangle=\mathbb{E}^A\left( u^{A,\lambda}\right)=\lambda^p\mathbb G^{A,\lambda}\left(\frac{u^{A,\lambda}}{\lambda}\right)=\lambda^p\mathbb G^{A,\lambda}\left(v^{A,\lambda}\right).
\end{equation}
Therefore, \Cref{Thm_conv_energy} implies that:
\begin{equation}
\label{limGBl_dim}
\lim_{\lambda\to +\infty} \mathbb G^{A,\lambda}\left(v^{A,\lambda}\right)=
\begin{dcases}
\mathbb B^A(z^A) &\text{if } p>q\\
\mathbb B^A(w^A) &\text{if } p<q\\
\mathbb B^A(v^A) &\text{if } p=q
\end{dcases}.
\end{equation}
where the Dirichlet Energy $\mathbb B^A$ has been defined in 
\eqref{Benergy}. In any of the three cases, the differential problem is Laplace modeled with $\gamma_b(x,s)=\beta(x)$, 
and
\begin{equation}
    \label{gamma_3_cases}
\gamma_a(x,s)=
\begin{cases}
0  &\text{if } p>q\\
+\infty&\text{if } p<q\\
\alpha(x)
 &\text{if } p=q\\
\end{cases}.
\end{equation}

Therefore, by \Cref{transferthm}, the energy terms in the r.h.s. of \eqref{limGBl_dim} are all equal to the associated A-DtN map, that particularizes as the DTN map over $p$, i.e.:
\begin{equation}
\label{3en=dl_dim}
\mathbb B^A(\cdot) =\frac 1p \langle\Lambda_\infty^A(f),f\rangle,
\end{equation}

Hence, by \eqref{chainLGl_dim}-\eqref{limGBl_dim}-\eqref{3en=dl_dim}, lead to the conclusion \eqref{lim_ADtN_DtN_dim}.
\end{proof}

\subsubsection{The $p_0-$Laplace Signature with small boundary data}
\label{subsecasslimsma}
The limiting case of problem \eqref{gminimum} for small Dirichlet boundary data is treated \cite{corboesposito2024thep0laplacesignature}, i.e. when the data is $\lambda f$ as $\lambda\to 0^+$.

In the limit of \lq\lq small'' boundary data, the existence of the solution for problem \eqref{problem} is guaranteed by requiring the following limiting behaviours for the background and the anomaly material properties.
\begin{itemize}
\item[{\bf (P$_0$3)}] There exist an exponent $p_0$ with $1< p_0 \leq p<\infty$ and a function $\beta_0$ such that:
    \begin{equation*}
        \begin{split}
            \lim_{s\to 0^+} \frac{\gamma_b (x,s)}{s^{p_0-2}}=\beta_0(x)\quad \text{for a.e.}\  x\in B.
        \end{split}
    \end{equation*}
\item[{\bf (Q$_0$3)}] There exist an exponent $q_0$ with $1< q_0 \leq q<\infty$ and a function $\alpha_0$ such that:
    \begin{equation*}
        \begin{split}
            \lim_{s\to 0^+} \frac{\gamma_a (x,s)}{s^{q_0-2}}=\alpha_0(x)\quad \text{for a.e.}\  x\in A.
        \end{split}
    \end{equation*}
\end{itemize}
Furthermore, the hypotheses {\bf (P1)}, {\bf(P2)}, {\bf (Q1)}, {\bf(Q2)} are complemented by the following ones, taking into account the role played by the exponent $p_0$ (see \cite{corboesposito2024thep0laplacesignature} for further details).

\begin{itemize}
\item[{\bf (P$_0$4)}] For fixed $p_0$ with $1< p_0 \leq p<\infty$, there exist two positive constants $\overline\gamma$ and $s_0$ such that: 
\[
\gamma_b(x, s)\leq\overline \gamma\max\left\{ \left(\frac{s}{s_0}\right)^{p_0-2}, \left(\frac{s}{s_0}\right)^{p-2}\right\}
\]
for a.e. $x\in B$ and for any $s\ge 0$.
\item[{\bf (P$_0$5)}] 
For fixed $p_0$ with $1< p_0 \leq p<\infty$, there exist two positive constants $\underline{\gamma}$ and $s_0$ such that:
\begin{equation*}
\left(\gamma_b(x,s_2)\frac{{\bf s}_2}{s_0}-\gamma_b(x,s_1)\frac{{\bf s}_1}{s_0}\right)\cdot     \left(\frac{{\bf s}_2}{s_0}-\frac{{\bf s}_1}{s_0}\right)\geq
        \begin{cases}
            \displaystyle\underline\gamma\left|\frac{{\bf s}_2}{s_0}-\frac{{\bf s}_1}{s_0}\right|^{p_0}\ &\text{if} \ p_0\geq 2\\ \displaystyle
            \underline{\gamma}\left(1+ \left|\frac{{\bf s}_2}{s_0}\right|^2+\left|\frac{{\bf s}_1}{s_0}\right|^2\right)^\frac{p_0-2}2\left|\frac{{\bf s}_2}{s_0}-\frac{{\bf s}_1}{s_0}\right|^2\ &\text{if}\ 1<p_0<2
        \end{cases}
\end{equation*}
for a.e. $x\in B$ and $\forall\ {\bf s}_1,{\bf s}_2\in\R^n$.
\item[{\bf (Q$_0$4)}] For fixed $q_0$ with $1< q_0 \leq q<\infty$, there exist two positive constants $\overline\gamma$ and $s_0$ such that:
\[
\gamma_a(x, s)\leq\overline \gamma\max\left\{  \left(\frac{s}{s_0}\right)^{q_0-2}, \left(\frac{s}{s_0}\right)^{q_0-2}\right\}
\]
for a.e. $x\in A$ and for any $s\ge 0$.
\item[{\bf (Q$_0$5)}] 
For fixed $q_0$ with $1< q_0 \leq p<\infty$, there exist two positive constant $\underline\gamma$ and $s_0$ such that:
\begin{equation*}
\left(\gamma_a(x,s_2)\frac{{\bf s}_2}{s_0}- \gamma_a(x,s_1)\frac{{\bf s}_1}{s_0}\right) \cdot\left( \frac{{\bf s}_2}{s_0}-\frac{{\bf s}_1}{s_0}\right)\geq
\begin{cases}\displaystyle
\underline\gamma\left|\frac{{\bf s}_2}{s_0}-\frac{{\bf s}_1}{s_0}\right|^{q_0}\ &\text{if} \ q_0\geq 2\\ \displaystyle
\underline\gamma\left(1+ \left|\frac{{\bf s}_2}{s_0}\right|^2+\left|\frac{{\bf s}_1}{s_0}\right|^2\right)^\frac{q_0-2}2\left|\frac{{\bf s}_2}{s_0}-\frac{{\bf s}_1}{s_0}\right|^2\ &\text{if}\ 1<q_0<2
        \end{cases}
\end{equation*}
for a.e. $x\in A$ and $\forall\ {\bf s}_1,{\bf s}_2\in\R^n$.
\end{itemize}

In this framework, the normalized solution $v_0^{A,\lambda}$ is introduced:
\begin{equation}
\label{reluv0}
v_0^{A,\lambda}=\frac{u^{A,\lambda}}{\lambda},
\end{equation}
where $u^{A,\lambda}$  is the minimizer of \eqref{gminimum} for a boundary data equal to $\lambda f$. Therefore $v^{A,\lambda}_0$ minimizes
\begin{equation}
\label{G0energy}
\begin{split}
&\qquad\qquad\qquad\qquad\min_{\substack{\xi\in W^{1,p_0}(\Omega)\cup W^{1,q_0}(\Omega)\\ \xi=f\ \text{on}\ \partial \Omega}}\mathbb G_0^{A,\lambda}(\xi),\\
&\mathbb G_0^{A,\lambda}(\xi)=\frac 1 {\lambda^{p_0}}\left(\int_{B} Q_b(x,\lambda|\nabla \xi(x)|)dx+\int_{A} Q_a(x,\lambda|\nabla \xi(x)|)dx\right).\end{split}
\end{equation}

Furthermore, the following relationship holds between the normalized and unnormalized Dirichlet Energy:
\begin{equation}
\label{relEG0}
\mathbb E^A(\xi)=\lambda^{p_0} \mathbb G_0^{A,\lambda}\left(\frac { \xi}{\lambda}\right)\qquad\forall \xi\in W^{1,p_0}(\Omega)\cup W^{1,q_0}(\Omega).
\end{equation}

To treat the asymptotic case, it is needed to introduce $w_0$, denoting the solution of problem \eqref{problem_PEC} for $\gamma_b(x,s)=\beta_0(x)s^{p_0-2}$, i.e. for an anomalous region filled by a PEC and a monomial-type nonlinearity in the background material; equivalently (see also the beginning of the Section 5 in \cite{corboesposito2024thep0laplacesignature}) $w_0$ is the minimizer of problem \eqref{minimum_PEC} when the energy $\mathbb E^A$ is replaced by
\begin{equation}
\label{B0energy} \mathbb B_0^A(\xi)=\int_{B} \beta_0(x)|\nabla \xi(x)|^{p_0} dx.
\end{equation}
Let us observe that $w_0$ is constant in any connected component of $A$. 

Whereas, $v_0^A$ denotes the solution of problem \eqref{problem_PEI} for $\gamma_b(x,s)=\beta_0(x)s^{p_0-2}$, i.e. for an anomalous region filled by a PEI and a monomial-type nonlinearity in the background material; equivalently  $v_0^A$ is the minimizer of problem \eqref{minimum_PEI} when the energy $\mathbb E^A$ is replaced by \eqref{B0energy}.

The limiting behaviours of the normalized Dirichlet Energy for small boundary data are the byproduct of \cite{corboesposito2024thep0laplacesignature}, formalized in (5.9) and (5.14), see also Remark 5.4.

\begin{theorem}
\label{Thm_conv_energy0}
Let $1<q_0<p_0<+\infty$ be such that $p_0\leq p$, $q_0\leq q$, $f\in X^{p}_\diamond(\partial \Omega)$, $\gamma^A$ satisfying {\bf (A)}, {\bf (P1)}, {\bf (P2)},  {\bf (P$_0$3)}, {\bf (P$_0$4)}, {\bf (P$_0$5)}, {\bf (Q1)}, {\bf (Q2)}, {\bf (Q$_0$3)}, {\bf (Q$_0$4)} and {\bf (Q$_0$5)}. Then
\begin{align}  
&\lim_{\lambda\to +\infty}\mathbb G_0^{A,\lambda}(v_0^{A,\lambda}) = \mathbb B_0^A(w_0^A)\quad \text{if } p_0>q_0,\\
&\lim_{\lambda\to +\infty}\mathbb G_0^{A,\lambda}(v_0^{A,\lambda}) = \mathbb B_0^A(z_0^A)\quad \text{if } p_0< q_0,\\
&\lim_{\lambda\to +\infty}\mathbb G^{A,\lambda}(v^{A,\lambda}) = \mathbb B^A(v_0^A)\quad \text{if } p_0= q_0,
\end{align}
where $v_0^{A,\lambda}\in W^{1,p}(\Omega)\cup W^{1,q}(\Omega)$ is the solution of \eqref{G0energy}, $w_0\in W^{1,p}(B)$ is the unique solution of  \eqref{minimum_PEC}; $z_0^A\in W^{1,p}(B)$
is the unique solution of
\eqref{minimum_PEI}, for $\gamma_b(x,s)=\beta_0(x)s^{p-2}$; and $v_0^A\in W^{1,p}(\Omega)$ is the unique solution of \eqref{P1}, for $\gamma_b(x,s)=\beta_0(x)s^{p-2}$, and $\gamma_a(x,s)=\alpha_0(x)s^{p-2}$.
\end{theorem}
In summary, it is observed that, when $q_0<p_0$, the conductor in region $A$ is replaced by a PEC, and in the region $B$ can reliably be modelled a $p_0-$Laplace problem, with a boundary condition given by a constant scalar potential $u$, on each connected component of $\partial A$.  Moreover, when $q_0<p_0$,  the limiting problem in 
$A$ must be assimilated to a PEI, and in $B$ is modeled by a $p_0-$Laplacian. Similarly, to the large boundary data setting, if $p_0=q_0$, the problem reduces to a $p_0$-Laplacian one, with at most a discontinuity in the coefficients.

At this stage, analogously to the large data regime, the previous convergence results for the energy are used to derive a new result establishing the corresponding convergence of the A-DtN maps in the small data regime.

\begin{theorem}
\label{thm_conv_aDTN0}
Let $1<q_0<p_0<+\infty$ be such that $p_0\leq p$, $q_0\leq q$, $f\in X^{p}_\diamond(\partial \Omega)$, $\gamma^A$ satisfying {\bf (A)}, {\bf (P1)}, {\bf (P2)},  {\bf (P$_0$3)}, {\bf (P$_0$4)}, {\bf (P$_0$5)}, {\bf (Q1)}, {\bf (Q2)}, {\bf (Q$_0$3)}, {\bf (Q$_0$4)} and {\bf (Q$_0$5)}. Then
\begin{equation}
\label{lim_ADtN_DtN_dim0}
\lim_{\lambda\to 0^+}\frac{\left\langle\overline{\Lambda}^A\left( \lambda f\right) ,\lambda f \right\rangle}{\lambda^{p_0}}=\frac 1{p_0} \langle\Lambda_0^A(f),f\rangle,
\end{equation}
where
\begin{equation}
\label{lambda_3_cases0}
\Lambda_0^A :f\in X^p_\diamond(\partial\Omega)\mapsto
\begin{cases}
\beta(x) \partial_n w_0^A|_{\partial\Omega} \in X^{p_0}_\diamond(\partial\Omega)'\quad &\text{if } p_0>q_0\\
 \beta(x) \partial_n z_0^A|_{\partial\Omega} \in X^{p_0}_\diamond(\partial\Omega)' &\text{if } p_0<q_0\\
\beta(x) \partial_n v_0^A|_{\partial\Omega} \in X_\diamond^{p_0}(\partial\Omega)'&\text{if } p_0=q_0\\
\end{cases},
\end{equation}
and $z_0^A\in W^{1,p_0}(B)$, $w_0^A\in W^{1,p_0}(B)$, and $v_0^A\in W^{1,p_0}(\Omega)$ are the unique solutions of \eqref{minimum_PEI}, \eqref{minimum_PEC}, and \eqref{P1}, respectively, with $p_0$ replacing $p$.
\end{theorem}

\section{The reconstruction methods for bounded background materials}
\label{sec_p=2_quasi}
In this section, the problem of reconstructing a nonlinear anomaly embedded in a bounded background material property ($p=2$) is investigated. In particular, given the Monotonicity Principle introduced in \Cref{theMP}, a converse result is provided in order to highlight the theoretical limit of an MP-based reconstruction method.

It is worth noting that the case of a quasilinear anomaly embedded in a linear background was investigated in \cite{mottola2025theinverseobstacle, mottola2026corrigendumtheinverseobstacle}, which provided a converse result for the MP, while the corresponding imaging method was developed in \cite{mottola2024imaging}.

\begin{figure}[htb]
    \centering
    \includegraphics[width=1\linewidth]{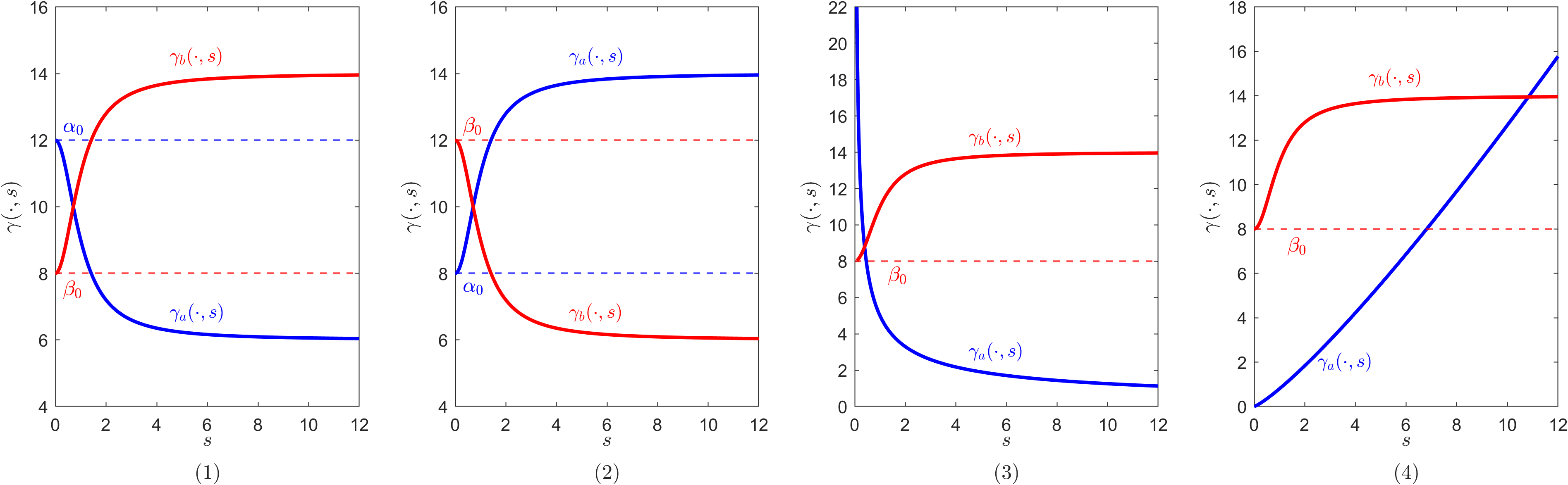}
    \caption{Examples of material properties corresponding to the four cases of \Cref{thm_quasi_pls}. The solid red line represents the linear background material property, and the solid blue line represents the nonlinear material property that fills the anomaly. Furthermore, the dashed lines correspond to the limiting behaviour $\beta_0$ and $\alpha_0$ of $\gamma_b(\cdot,s)$ and $\gamma_a(\cdot,s)$ as $s\to 0^+$, respectively.}
    \label{fig_04_quasi_p2}
\end{figure}

In this context, the goal of this section is to extend the analysis beyond linear backgrounds. Specifically, a genuinely quasilinear but bounded background material is investigated (see \Cref{fig_04_quasi_p2} for illustrative examples).

Before stating the main results of this section, the concept of outer support is recalled. Specifically, the outer support~\cite{harrach2013monotonicity,mottola2025theinverseobstacle,mottola2026corrigendumtheinverseobstacle} of a set $A\subseteq \Omega$ is defined as follows.
\begin{definition}\label{def:outer}
The outer support of a set $A \subseteq \Omega$, denoted as $A^*$, is the complement in $\overline{\Omega}$, of the union of those relatively open set $U$ contained in $\Omega \setminus\overline A$ and connected to $\partial \Omega$, i.e. those sets $U$ that are connected and satisfying $\partial U\cap\partial\Omega \neq \emptyset$. 
\end{definition}

The key idea for extending the results of the previous section to this more general setting is to properly exploit the knowledge of the limiting behaviour of the Dirichlet Energy, for either large or small boundary data, provided in \Cref{sec_p_sig}.

In the case of material properties fulfilling the asymptotic behaviours {\bf (P$_\infty$3)} and {\bf (Q$_\infty$3)}, the following converse result holds.
\begin{theorem}[Large boundary data]\label{thm_inf_conv}
Let $1<q<+\infty$, $f\in X^2_\diamond(\partial \Omega)$, $\gamma_b\in L_+^\infty (\Omega)$ be piecewise analytic, $\gamma^A$ and $\gamma^T$ defined as in \eqref{gammaA} satisfying {\bf (A)}, {\bf (P$_\infty$3)} and {\bf (Q$_\infty$3)}
and one of the following holds
\begin{enumerate}
\item $\gamma_a(x,s)\in L_{+}^{\infty}(\Omega)$, and $\alpha(x)>\beta(x)$ a.e. in $x\in\Omega$.
\item $\gamma_a(x,s)\in L_{+}^{\infty}(\Omega)$, and $\alpha(x)<\beta(x)$ a.e. in $x\in\Omega$.
\item $\gamma_a$ satisfying {\bf (Q1)} and {\bf (Q2)} both when $q > 2$.
\item $\gamma_a$ satisfying {\bf (Q1)} and {\bf (Q2)} both when $1 < q < 2$.
\end{enumerate}
Then, in the cases (1) or (3) (respectively (2) or (4)) it holds
\begin{equation}
\label{converse_negli}
T \nsubseteq A^* \Longrightarrow \overline{\Lambda}^T\nleqslant\overline{\Lambda}^A \  (\text{respectively }   \overline{\Lambda}^T\ngeqslant\overline{\Lambda}^A)\qquad \forall A, T\in\mathcal S (\Omega).
\end{equation}
Moreover, if $A \in \mathcal S (\Omega)$ has a connected complement, then
\begin{equation}
\label{doppia_implicazioneli}
T\subseteq A \iff \overline\Lambda^T\leqslant{\overline\Lambda}^A \ (\text{respectively } \overline\Lambda^T\geqslant{\overline\Lambda}^A) \qquad\forall\ T\in\mathcal S (\Omega).
\end{equation}
\end{theorem}
\begin{proof}
By \Cref{thm_conv_aDTN}, specialized to the case $p=2$, it holds:
\begin{equation}
    \label{lim_ADtN_DtN}
\lim_{\lambda\to +\infty}\frac{\left\langle\overline{\Lambda}^A\left( \lambda f\right) ,\lambda f \right\rangle}{\lambda^2}=\frac 12 \langle\Lambda_\infty^A(f),f\rangle.
\end{equation}

Regarding the cases (1) and (3), if $T\not\subseteq A^*$, there exists a sequence of boundary data $\{f_n\}_n\subset X_{\diamond}^2(\partial\Omega)$ (see \cite{harrach2013monotonicity}, \cite{candiani2019monotonicity} for boundary data modeled by the Neumann-to-Dirichlet operator, and \cite{mottola2025theinverseobstacle,mottola2026corrigendumtheinverseobstacle} for transferring the same results to the DtN operator), such that
\begin{equation*}
    \lim_{n\to+\infty} \langle (\Lambda_{\infty}^T-\Lambda_{\infty}^A)f_n,f_n \rangle =-\infty.
\end{equation*}
For any fixed $M>0$, there exists $m\in\mathbb{N}$ such that $
   \langle (\Lambda_{\infty}^A-\Lambda_{\infty}^T)f_m,f_m \rangle <-M
$, passing to the limit as $\lambda\to+\infty$, by using \eqref{lim_ADtN_DtN}, it holds
\begin{equation}
    \label{lim_together}
\lim_{\lambda\to +\infty}\frac{\left\langle\overline{\Lambda}^A\left( \lambda f_m\right)-\overline{\Lambda}^T\left( \lambda f_m\right) ,\lambda f_m \right\rangle}{\lambda^2}=\frac 12 \langle\Lambda_\infty^A(f_m)-\Lambda_\infty^T(f_m),f_m\rangle<-\frac M 2,
\end{equation}
that it is the conclusion
\eqref{converse_negli}.
Cases (2) and (4) follow by analogous arguments.

Finally, when the outer support of $A$ coincides with $A$, i.e. $A^*=A$, the monotonicity principle (stated at equation \eqref{m_charge} of \Cref{monothm})
gives the equivalence stated in \eqref{doppia_implicazioneli}.
\end{proof}

In the case of material properties fulfilling the limiting behaviours {\bf (P$_0$3)} and {\bf (Q$_0$3)}, for small boundary data, the analogous converse monotonicity result can be proved by following the same arguments as in the previous theorem, with \Cref{thm_conv_aDTN0} replacing \Cref{thm_conv_aDTN}.
\begin{theorem}[Small boundary data]\label{thm_quasi_pls}
Let $1<q_0<2<+\infty$ be such that $2\leq p$, $q_0\leq q$, $f\in X^{2}_\diamond(\partial \Omega)$, $\gamma_b\in L^\infty_+(\Omega)$ be piecewise analytic, $\gamma^A$ and $\gamma^T$ defined as in \eqref{gammaA} satisfying {\bf (A)}, {\bf (P1)}, {\bf (P2)}, {\bf (P$_0$3)}, {\bf (Q1)}, {\bf (Q2)}, {\bf (Q$_0$3)} and one of the following holds
\begin{enumerate}
\item $\gamma_a(x,s)\in L_{+}^{\infty}(\Omega)$, and $\alpha_0(x)>\beta_0(x)$ a.e. in $x\in\Omega$.
\item $\gamma_a(x,s)\in L_{+}^{\infty}(\Omega)$, and $\alpha_0(x)<\beta_0(x)$ a.e. in $x\in\Omega$.
\item $\gamma_a$ satisfying {\bf (Q$_0$4)} and {\bf (Q$_0$5)} both when $q_0 > 2$.
\item $\gamma_a$ satisfying {\bf (Q$_0$4)} and {\bf (Q$_0$5)} both when $1 < q_0 < 2$.
\end{enumerate}
Then, in the cases (1) or (3) (respectively (2) or (4)) it holds
\begin{equation}
\label{converse_negli0}
T \nsubseteq A^* \Longrightarrow \overline{\Lambda}^T\nleqslant\overline{\Lambda}^A \  (\text{respectively }   \overline{\Lambda}^T\ngeqslant\overline{\Lambda}^A)\qquad \forall A, T\in\mathcal S (\Omega).
\end{equation}
Moreover, if $A \in \mathcal S (\Omega)$ has a connected complement, then
\begin{equation}
\label{doppia_implicazioneli0}
T\subseteq A \iff \overline\Lambda^T\leqslant{\overline\Lambda}^A \ (\text{respectively } \overline\Lambda^T\geqslant{\overline\Lambda}^A) \qquad\forall\ T\in\mathcal S (\Omega).
\end{equation}
\end{theorem}

These results establish a converse monotonicity principle for generic quasilinear materials, provided that the background property is bounded and the asymptotic behaviour assumptions hold for either large or small boundary data. In particular, \Cref{thm_inf_conv,thm_quasi_pls} prove that, for noise-free data, an MP-based approach yields a reconstruction that never exceeds the outer support of the actual anomaly, even in the presence of a nonlinear background and a generic quasilinear anomaly.

Moreover, the results also cover cases in which the background and anomaly properties may intersect. Indeed, the order relation between the corresponding material properties with respect to $s$ need not hold for all the values of $s>0$; it is enough that it holds asymptotically in at least one of the two limiting regimes, either as $s\to 0^+$ or as $s\to+\infty$.

\section{The reconstruction methods for possibly unbounded or vanishing background material}\label{sec_p_methods}

In this Section, the problem of the obstacle reconstruction in the presence of a material property with growth exponent  $p\neq2$ is studied. Analogously to the results of the previous section, the inner region has a quasilinear behaviour with $1<q<+\infty$ growth.

Some preliminary geometrical definitions and results are recalled before stating the main Theorems of this section. From \cite[Definitions 2.4 and 2.5]{brander2018monotonicity}, the concept of the convex hull of a set is introduced.
\begin{definition}[closed essential convex hull]
For any $\rho\in\R^n$ and $t\in\R$, the closed half-space
is defined as
\[
H_{\rho, t} := \{ x \in \mathbb{R}^n : x \cdot \rho \leq t \}.
\]The (closed essential) convex hull $\mathrm{conv}\,(A)$ of a measurable set $A \subset \mathbb{R}^n$, is
\[
\bigcap_{\substack{\rho \in \mathbb{R}^n,\ t \in \mathbb{R}\\ m\bigl(A \setminus H_{\rho, t}\bigr) = 0}} H_{\rho, t}
\]
\end{definition}
\begin{definition}[essential convex support function]
The essential convex support function of a bounded measurable $A \subseteq \mathbb{R}^n$ is 
 $h_A : \mathbb S^{n-1} \rightarrow \mathbb{R}$ defined by
\[
h_A(\rho) = \inf \{ t \in \mathbb{R} : m(A \setminus H_{\rho, t}) = 0 \}.
\]  
\end{definition}

A key tool in proving a converse monotonicity principle is the existence of proper boundary excitations that make the gradient of the corresponding solution vanish in a certain subset of the domain, while ensuring a non-zero gradient in another region. In this setting, such boundary data are constructed starting from the so-called Wolff solutions \cite{wolff2007gap}, which are particular solutions to the problem 
\begin{equation}\label{eqn_p_prob}
\begin{cases}
\dive(\gamma_o |\nabla u(x)|^{p-2}\nabla u(x))=0 & \text{in $\Omega$}, \\
    u(x)=f(x) & \text{on $\partial\Omega$}
\end{cases}
\end{equation}
which decrease exponentially in a prescribed direction. This, in turn, allows to construct boundary potentials having the desired property, i.e., (i) the corresponding solution vanishes in a certain set, and (ii) the gradient remains well separated from zero in a different region, provided the two sets can be separated by a hyperplane.

In particular, the Wolff-type solutions are recalled in the following Lemma (see \cite{brander2018monotonicity,salozhong2012}).
\begin{lemma}
    Given $\mathbf{d}, \mathbf{d}^{\perp}\in \mathbb{R}^n$, such that $\mathbf{d}\cdot\mathbf{d}^{\perp}=0$, and $\tau,t\in\mathbb{R}$, the potential
    \begin{equation}\label{wolff_sol_bound}
        u(x)=e^{\tau (\mathbf{x}\cdot\mathbf{d}-t)}w(\tau \mathbf{x}\cdot\mathbf{d}^{\perp})
    \end{equation}
    is a solution of $\dive(|\nabla u|^{p-2}\nabla u)=0$ in $\mathbb{R}^n$.

    In the above expression, $w(x)$ is the solution of the following differential equation
    \begin{equation}\label{wolff_sol_1}
        w''(s) + V(w,w')w = 0,
    \end{equation}
    with
    \begin{equation}
        V(a,b) = \frac{(2p-3)b^2 + (p-1)a^2}{(p-1)b^2 + a^2},
    \end{equation}
    and initial conditions $(a_0,b_0)\neq(0,0)$.
\end{lemma}
The above solutions allow to define proper depleting potentials, as specified in the following result
\begin{proposition}
    \label{lem_wolff}
    Let $S_1, S_2$ be two measurable set, such that $S_2\not\subseteq\mathrm{conv}\, (S_1) = \varnothing$. Then, there exists a sequence of boundary data $\{f_n\}_{n\in\N} \subseteq X_\diamond^p(\partial\Omega)$ such that
    \begin{equation*}
        \lim_{n\to +\infty} \int_{\mathrm{conv}\, (S_1)} \gamma_0 |\nabla u_n(x)|^p\,dx=0 \quad \text{and} \quad  \lim_{n\to +\infty} \int_{S_2} \gamma_0 |\nabla u_n(x)|^p\,dx=+\infty,
    \end{equation*}
    where $u_n$ is the solution of the problem \eqref{eqn_p_prob} for $f=f_n$.
\end{proposition}
\begin{proof}
To prove the claim, it is enough to show that the ratio between the energy
concentrated in the $S_2$  and the energy concentrated in $conv (S_1)$ diverges to $+\infty$. Indeed, the operator mapping the boundary data into the solution of problem \eqref{eqn_p_prob} is homogeneous of degree 1, i.e. the solution corresponding to $k f$ is $k u$, it possible to properly normalize the sequence $f_n$, in order to obtain the claim.

    By standard arguments (see for example \cite{brander2018monotonicity}), it is possible to show that there exists a point $x_0\in S_2$, and a sufficiently small $r>0$, such that the open ball $B_r(x_0)\subset S_2$ has positive measure and $B_r(x_0) \cap \mathrm{conv}\,(S_1) =\varnothing$. Furthermore, there exists $\mathbf{d}\in\mathbb{R}^n$, $\mathbf{d}=1$, and $t\in\mathbb{R}$, such that
    \begin{alignat*} {3}
        &\mathbf{x}\cdot\mathbf{d}&&<t \quad&& \forall\, \mathbf x\in \mathrm{conv}\,(S_1),  \\
        &\mathbf{x}\cdot\mathbf{d}&&>t+\varepsilon \quad&& \forall\, \mathbf x\in B_r(x_0). 
    \end{alignat*}
    Considering the Wolff solutions introduced in Lemma \ref{lem_wolff}, and by \cite[Lemma 2.8]{brander2018monotonicity}, there exists $c,C>0$, such that for each $\tau>0$, there exists a solution $u_{\tau}$ of problem \eqref{eqn_p_prob} fulfilling
    \begin{equation*}
        c \tau e^{\tau(x\cdot d-t)}\leq |\nabla u_{\tau} (x)|\leq C \tau e^{\tau(x\cdot d-t)}, \quad \forall\, x \in \Omega.
    \end{equation*}

    Let $f_n=u_{\tau_n}|_{\partial\Omega}$, where $\tau_n>1$ is a monotonically increasing sequence, and let $u_n$ the corresponding solution of problem \eqref{eqn_p_prob}, it holds
    \begin{equation*}
        \int_{B} \gamma_0 |\nabla u_n(x)|^p\,dx \geq \int_{B_r} \gamma_0 |\nabla u_n(x)|^p\,dx \geq \gamma_0 c^p\tau_n^p e^{p\tau_n\varepsilon} |B_r|^p,
    \end{equation*}
    while 
    \begin{equation*}
        \int_{\mathrm{conv} (S_1)} \gamma_0 |\nabla u_n(x)|^p\,dx \leq \gamma_0 C^p \tau_n^p |\mathrm{conv} (S_1)|^p,
    \end{equation*}
    where $|B_r|$, $|\mathrm{conv} (S_1)|$, are the measures of the sets $B_r$ and $\mathrm{conv} (S_1)$, respectively.

    As a consequence,
    \begin{equation*}
        \lim_{n\to +\infty} \frac{\displaystyle\int_{B} \gamma_0 |\nabla u_n(x)|^p\,dx}{\displaystyle\int_{\mathrm{conv} (S_1)} \gamma_0 |\nabla u_n(x)|^p\,dx}=+\infty,
    \end{equation*}
    that concludes.
\end{proof}

\subsection{The $p-$Laplace case}
\label{pnot2_Lap}
\begin{figure}[htb]
    \centering
    \includegraphics[width=1\linewidth]{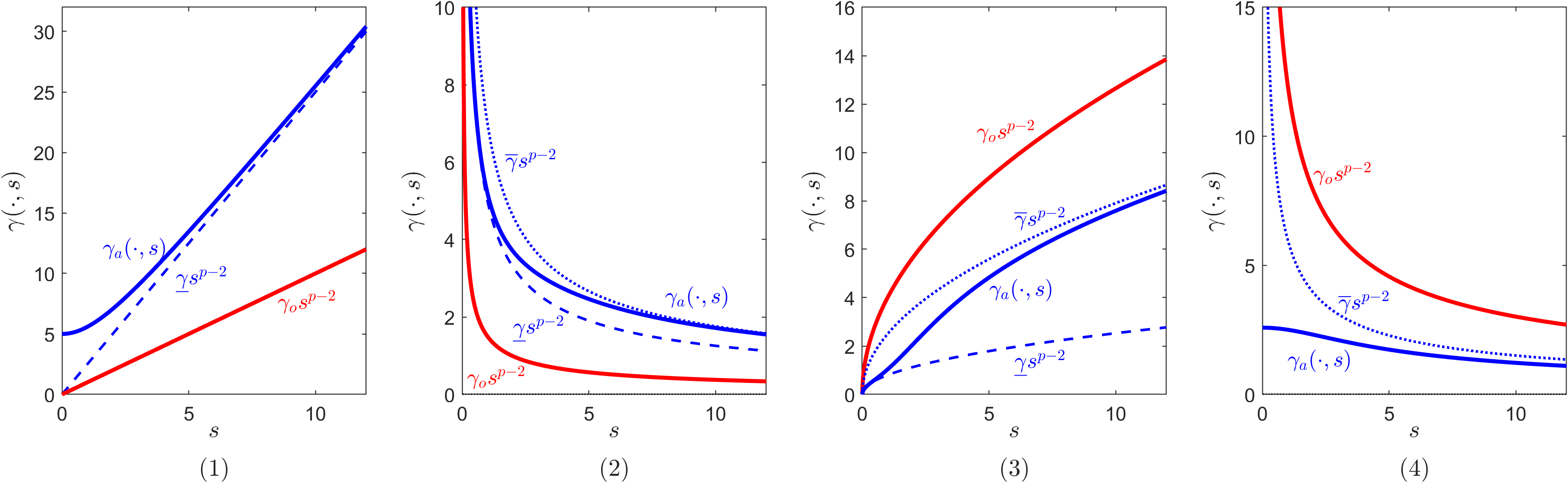}
    \caption{Examples of material properties fulfilling the hypothesis of \Cref{thm_converse_p_esteso}.}
    \label{fig_05_lin_pn2}
\end{figure}

In this section, the background material property is assumed to have $p-$Laplacian behaviour, i.e.
\begin{equation*}
    \gamma_b(x,s)=\gamma_o s^{p-2},
\end{equation*}
where $\gamma_o$ is a positive constant, while the material property for the anomaly is allowed to be quasilinear, with the same growth exponent $q=p$ (see \Cref{fig_05_lin_pn2} for some examples). It is worth noticing that although the case $p=2$ is covered by the following theorem, the results of the previous section provide a simpler approach.

To be more specific, the following material properties are considered.
\begin{equation}
\label{gammalinAp}
\gamma^A(x,s)=\begin{cases}
   \gamma_o s^{p-2} & \text {in $\Omega\setminus A$} \\
   \gamma_{a}(x,s) & \text {in $A$}
  \end{cases},
\end{equation}
and
\begin{equation}
\gamma^T(x)=
\begin{cases}
\gamma_o s^{p-2} & \text{in $\Omega\setminus T$}\\
\underline{\gamma} s^{p-2} & \text{in $T$}
\end{cases},
\label{eqn:sigmaTp1}
\end{equation}
or
\begin{equation}
\gamma^T(x)=
\begin{cases}
\gamma_o s^{p-2} & \text{in $\Omega\setminus T$}\\
\overline{\gamma} s^{p-2} & \text{in $T$}
\end{cases}.
\label{eqn:sigmaTp2}
\end{equation}

   
The following theorem establishes a (partial) converse result for the A-MPM in the presence of a $p$-Laplacian background and a generic quasilinear anomaly, under the condition that $p=q$. Specifically, leveraging Wolff solutions makes it possible to show that, in this nonlinear setting, an MP-based imaging method yields a reconstruction that never exceeds the convex hull of the actual anomaly.
\begin{theorem}
\label{thm_converse_p_esteso}
Let $1<p=q<+\infty$, $\gamma_o$ be a positive constant, $\gamma^A(x,s)$ be defined in \eqref{gammalinAp} satisfying {\bf (A)}, {\bf (P1)}, {\bf (P2)}, and one of the following holds
\begin{enumerate}
\item $\gamma_a$ satisfying {\bf (Q1)} and {\bf (Q2)} for $p > 2$ and $\gamma_o < \underline{\gamma}$, and  the (test) material property $\gamma^T$ is defined as \eqref{eqn:sigmaTp1}.
\item $\gamma_a$ satisfying {\bf (Q1)} for $1< p < 2$, $\gamma_a(x,s)\geq \underline{\gamma}\left(\frac{s}{s_0}\right)^{p-2}$, $\gamma_o < \underline{\gamma}$, and  the (test) material property $\gamma^T$ is defined as \eqref{eqn:sigmaTp1}.
\item $\gamma_a$ satisfying {\bf (Q2)} for $p > 2$, $\gamma_a(x,s)\leq \overline{\gamma}\left(\frac{s}{s_0}\right)^{p-2}$, $\gamma_o > \overline{\gamma}$, and  the (test) material property $\gamma^T$ is defined as \eqref{eqn:sigmaTp2}.
\item $\gamma_a$ satisfying {\bf (Q1)} and {\bf (Q2)} for $1 < p < 2$ and $\gamma_o > \overline{\gamma}$, and  the (test) material property $\gamma^T$ is defined as \eqref{eqn:sigmaTp2}.
\end{enumerate}
Then, in the case (1) or (2) 
(respectively (3) or (4)) it holds
\begin{equation}
\label{converse_neg_casop}
T \nsubseteq \mathrm{conv}\,(A) \Longrightarrow \overline{\Lambda}^T\nleqslant\overline{\Lambda}^A \  (\text{respectively }   \overline{\Lambda}^T\ngeqslant\overline{\Lambda}^A)\qquad \forall A, T\in\mathcal S (\Omega).
\end{equation}
Moreover, if $A \in \mathcal S (\Omega)$ is convex, then
\begin{equation}
\label{doppia_implicazione_casop}
T\subseteq A \iff \overline\Lambda^T\leqslant{\overline\Lambda}^A \ (\text{respectively } \overline\Lambda^T\geqslant{\overline\Lambda}^A) \qquad\forall\ T\in\mathcal S (\Omega).
\end{equation}
\end{theorem}
\begin{proof}
\underline{\it Case (1)-(2).}  Preliminarily, it is observed that
\begin{equation}\label{eqn_p_split}
    \langle \overline{\Lambda}^A(f)-\overline{\Lambda}^T(f),f\rangle=\langle \overline{\Lambda}^A(f)-\overline{\Lambda}^{\varnothing}(f),f\rangle-\langle \overline{\Lambda}^T(f)-\overline{\Lambda}^{\varnothing}(f),f\rangle,
\end{equation}
where $\overline{\Lambda}^{\varnothing}$ is the A-DtN operator corresponding to the material property $\gamma^{\varnothing}(x,s)=\gamma_o s^{p-2}$, for a.e $x\in\Omega$.
Denoting by $u^{\varnothing}$ and $u^A$, the solutions of problem \eqref{problem} when the material property is $\gamma^{\varnothing}$ and $\gamma^A$, respectively, it follows
\begin{equation}\label{eqn_ub_1}
\begin{split}
\langle \overline{\Lambda}^A(f)-\overline{\Lambda}^{\varnothing}(f),f\rangle&\leq \int_{\Omega}\int_0^{\abs{\nabla u^{\varnothing}}}(\gamma^A(x,\xi)-\gamma^{\varnothing}(x,\xi))\xi\,d\xi \\
&=\int_{A}\int_0^{\abs{\nabla u^{\varnothing}}}(\gamma_a(x,\xi)-\gamma_o\xi^{p-2})\xi\,d\xi \\
&\leq \begin{cases}
\displaystyle\int_{A}\int_0^{\abs{\nabla u^{\varnothing}}}\left[\overline{\gamma}\xi^{p-2}-\gamma_o\xi^{p-2}\right]\xi\,d\xi &\text{ if } 1<p<2 \\ \displaystyle\int_{A}\int_0^{\abs{\nabla u^{\varnothing}}}\left[\overline{\gamma}\left(1+\xi^{p-2}\right)-\gamma_o\xi^{p-2}\right]\xi\,d\xi &\text{ if } p>2\end{cases}\\
& =\begin{cases}
\displaystyle \frac{1}{p}\int_A (\overline{\gamma}-\gamma_o)\abs{\nabla u^{\varnothing}}^p(x)\,dx&\text{ if } 1<p<2\\    
\displaystyle \frac{1}{2}\int_A \overline{\gamma}\abs{\nabla u^{\varnothing}}^2(x)\,dx+\frac{1}{p}\int_A (\overline{\gamma}-\gamma_o)\abs{\nabla u^{\varnothing}}^p(x)\,dx&\text{ if } p>2,
  \end{cases}  \end{split}
\end{equation}
where the first line follows from Theorem \ref{transferthm} and the minimality of the Dirichlet Energy; the second line follows from $\gamma^A$ and $\gamma^{\varnothing}$ agreeing on $\Omega\setminus A$; and the third line follows from {\bf (Q1)}, for $s_0=1$.

Moreover, a lower bound can be established for the second term in \eqref{eqn_p_split}, as follows
\begin{equation}\label{eqn_lb_1}
\begin{split}
    \langle \overline{\Lambda}^T(f)-\overline{\Lambda}^{\varnothing}(f),f\rangle &= \frac{1}{p}\langle \Lambda^T(f)-\Lambda^{\varnothing}(f),f\rangle \\    
    & \geq \frac{p-1}{p} \int_{T} \frac{\gamma^{\varnothing}}{\underline{\gamma}^{\frac{1}{p-1}}}\left(\underline{\gamma}^{\frac{1}{p-1}}-\gamma_o^{\frac{1}{p-1}}\right) \abs{\nabla u^{\varnothing}}^p\,dx,
\end{split} 
\end{equation}
where the first line follows from the definition of Dirichlet Energy, while the second line follows from \cite[Lemma 2.9]{brander2018monotonicity} and the fact that $\gamma^T$ and $\gamma^{\varnothing}$ agree on $\Omega\setminus T$.

Let $T\not\subseteq \textrm{conv}\, (A)$, by \Cref{lem_wolff}, for $S_1=\mathrm{conv}\,(A)$ and $S_2=T$, there exists a sequence of boundary data $\{f_n\}_{n\in\N}$ and a corresponding sequence $\{u_n\}_{n\in\N}$ of solutions for problem \eqref{eqn_p_prob}, such that 
\begin{align}
\label{eqn_w_1}
    \lim_{n\to +\infty} \int_{\textrm{conv}(A)} \gamma_o \abs{\nabla u_n}^p(x)\,dx&=0, \\
    \label{eqn_w_2}
    \lim_{n\to +\infty} \int_T \gamma_o \abs{\nabla u_n}^p(x)\,dx&=+\infty.
\end{align}
By combining \eqref{eqn_p_split}, \eqref{eqn_ub_1}, \eqref{eqn_lb_1}, particularized for $f=f_n$ and $u^{\varnothing}=u_n$: 
\begin{equation}\label{p_estimate_1}
\begin{split}
        \langle \overline{\Lambda}^A(f_n)-\overline{\Lambda}^T(f_n),f_n\rangle&\leq \frac{1}{2}\int_A \overline{\gamma}\abs{\nabla u_n}^2(x)\,dx+\frac{1}{p}\int_A (\overline{\gamma}-\gamma_o)\abs{\nabla u_n}^p(x)\,dx\\
        &\quad-  \frac{p-1}{p} \int_{T} \frac{\gamma^{\varnothing}}{\underline{\gamma}^{\frac{1}{p-1}}}\left(\underline{\gamma}^{\frac{1}{p-1}}-\gamma_o^{\frac{1}{p-1}}\right) \abs{\nabla u_n}^p\,dx.
\end{split}
\end{equation}
Observing that $A \subseteq \mathrm{conv}(A)$, and noticing that the 
convergence of $\|\nabla u_n\|_{L^p(A)}$ to zero implies the convergence
of $\|\nabla u_n\|_{L^2(A)}$ to zero, and since $A$ is bounded, then by using 
\eqref{eqn_w_1}-\eqref{eqn_w_2} in \Cref{p_estimate_1}, the equation \eqref{converse_neg_casop} follows.

Moreover, assumptions $\gamma_0<\underline{\gamma}$, and $\gamma_a(x,s)\geq \underline{\gamma}s^{p-2}$ for $1<p<2$, ensures that $\gamma^A(x,s)\geq \gamma^T(x,s)$ for a.e. $x\in\Omega$ and $s>0$, when $T\subseteq A$, i.e. that
$    T\subseteq A \Longrightarrow \overline{\Lambda}^T\leqslant\overline{\Lambda}^A,$
giving \eqref{doppia_implicazione_casop}.

\underline{\it Case (3)-(4)}. The following lower bound on $\overline\Lambda^A$ will be useful:
\begin{equation}
\label{chain_Lambda_A}
\begin{split}
\langle \overline{\Lambda}^A(f),f\rangle &= \int_{\Omega}\int_0^{\abs{\nabla u^A}} \gamma^A(x,\xi)\xi\,d\xi dx\\ 
&\geq
\begin{cases}
    \displaystyle \int_{\Omega\setminus A}\int_0^{\abs{\nabla u^A}} \gamma_o \xi^{p-2}\xi\,d\xi dx+ \int_{A}\int_0^{\abs{\nabla u^A}} \underline{\gamma} (1+\xi^2)^{(p-2)/2}\xi\,d\xi dx & \text{for } 1<p<2\\
    \displaystyle \int_{\Omega\setminus A}\int_0^{\abs{\nabla u^A}} \gamma_o \xi^{p-2}\xi\,d\xi dx+ \int_{A}\int_0^{\abs{\nabla u^A}} \underline{\gamma} \xi^{p-2}\xi\,d\xi dx & \text{for } p>2\\
\end{cases}\\
&= 
\begin{cases}
    \displaystyle \frac{1}{p}\int_{\Omega\setminus A} \gamma_o \abs{\nabla u^A}^p\,dx + \frac{1}{p}\int_{A} \underline{\gamma}\left(\left(1+\abs{\nabla u^A}^2\right)^{p/2}-1\right)\,dx & \text{for } 1<p<2 \\ 
    \displaystyle \frac{1}{p}\int_{\Omega\setminus A} \gamma_o \abs{\nabla u^A}^p\,dx + \frac{1}{p}\int_{\Omega\setminus A} \underline{\gamma} \abs{\nabla u^A}^p\,dx & \text{for } p>2 \\ 
\end{cases}\\
&\begin{cases}
   \displaystyle \geq \frac{1}{p}\left[\int_{\Omega\setminus A} \gamma_o \abs{\nabla u^A}^p\,dx + \int_{A} \underline{\gamma} \abs{\nabla u^A}^p\,dx - \underline{\gamma}\abs{A}\right] & \text{for } 1<p<2\\  
    \displaystyle = \frac{1}{p}\int_{\Omega\setminus A} \gamma_o \abs{\nabla u^A}^p\,dx + \int_{A} \underline{\gamma} \abs{\nabla u^A}^p\,dx & \text{for } p>2\\ 
\end{cases}\\
& = 
\begin{cases}
    \displaystyle\int_{\Omega}\int_0^{\abs{\nabla u^A}} \gamma^{A,l}(x,\xi)\,d\xi dx -\frac{\underline{\gamma}}{p}\abs{A} & \text{for } 1<p<2 \\
    \displaystyle\int_{\Omega}\int_0^{\abs{\nabla u^A}} \gamma^{A,l}(x,\xi)\,d\xi dx  & \text{for } p>2 
\end{cases} \\
& = 
\begin{cases}
    \displaystyle \langle \overline{\Lambda}^{A,l}(f),f\rangle - \frac{\underline{\gamma}}{p}\abs{A} & \text{for } 1<p<2 \\
    \displaystyle \langle \overline{\Lambda}^{A,l}(f),f\rangle  & \text{for } p>2
\end{cases}
\end{split}
\end{equation}
where the first line follows by \Cref{transferthm}, the second line follows by assumption {\bf{(Q2)}}, in the third line the inner integral has been computed, in the fourth line it has been exploited that $\left(1+\abs{\nabla u^A}^2\right)^{p/2}\geq \abs{\nabla u^A}^p$ for $1<p<2$, the fifth line follows by introducing the material property $\gamma^{A,l}$, defined as
\begin{equation*}
        \gamma^{A,l}(x,s) = \begin{cases}
            \gamma_o s^{p-2} & \text{in } \Omega\setminus A \\
        \underline{\gamma} s^{p-2} & \text{in }A,
        \end{cases}
\end{equation*}
with the related DtN operator $\Lambda^{A,l}$; and the last line follows again by \Cref{transferthm}.

Let $T\not\subseteq \textrm{conv}\, (A)$, analogously to case {\it (1)-(2)}, it is observed that
\begin{equation}\label{eqn_p_split_2}
        \langle \overline{\Lambda}^T(f)-\overline{\Lambda}^A(f),f\rangle 
        =\langle \overline{\Lambda}^T(f)-\overline{\Lambda}^{\varnothing}(f),f\rangle-\langle \overline{\Lambda}^A(f)-\overline{\Lambda}^{\varnothing}(f),f\rangle.
    \end{equation}

The first term at the r.h.s. can be treated using from Theorem \ref{transferthm} and the minimality of the Dirichlet Energy to gain:
\begin{equation}\label{eqn_ub_lb_2}
            \langle \overline{\Lambda}^T(f)-\overline{\Lambda}^{\varnothing}(f),f\rangle \leq  \frac{1}{p}\int_T (\overline{\gamma}-\gamma_o)\abs{\nabla u^{\varnothing}}^p(x)\,dx.
    \end{equation}
Regading the second term, by using the equation \eqref{chain_Lambda_A} and \cite[Lemma 2.9]{brander2018monotonicity}, it holds
\begin{equation}
\label{bound_l_34}
    \langle \overline{\Lambda}^A(f)-\overline{\Lambda}^{\varnothing}(f),f\rangle \geq 
    \begin{cases}
\displaystyle\frac{p-1}{p} \int_{A} \frac{\gamma_o}{\underline{\gamma}^{\frac{1}{p-1}}}\left(\underline{\gamma}^{\frac{1}{p-1}}-\gamma_o^{\frac{1}{p-1}}\right) \abs{\nabla u^{\varnothing}}^p\,dx-\frac{\underline{\gamma}}{p}\abs{A} & \text{for } 1<p<2 \\
        \displaystyle\frac{p-1}{p} \int_{A} \frac{\gamma_o}{\underline{\gamma}^{\frac{1}{p-1}}}\left(\underline{\gamma}^{\frac{1}{p-1}}-\gamma_o^{\frac{1}{p-1}}\right) \abs{\nabla u^{\varnothing}}^p\,dx & \text{for } p>2,
    \end{cases}
\end{equation}

Combining \eqref{eqn_p_split_2}, \eqref{eqn_ub_lb_2} and \eqref{bound_l_34}, for $f=f_n$ and $u^{\varnothing}=u_n$, as defined in \Cref{lem_wolff} in the setting $S_1=\textrm{conv}(A)$ and $S_2=T$, it holds
\begin{equation}\label{p_estimate_2}
\begin{split}
    \langle \overline{\Lambda}^T(f_n)-\overline{\Lambda}^A(f_n),f_n\rangle &\leq \frac{1}{p}\int_T (\overline{\gamma}-\gamma_o)\abs{\nabla u_n}^p(x)\,dx  \\
    & \quad -\frac{p-1}{p} \int_{A} \frac{\gamma_o}{\underline{\gamma}^{\frac{1}{p-1}}}\left(\underline{\gamma}^{\frac{1}{p-1}}-\gamma_o^{\frac{1}{p-1}}\right) \abs{\nabla u_n}^p\,dx+\frac{\underline{\gamma}}{p}\abs{A} 
\end{split}
\end{equation}
\Cref{converse_neg_casop} follows by noticing that $\overline{\gamma}-\gamma_o<0$.

Moreover, assumptions $\gamma_o>\overline{\gamma}$, and $\gamma_a(x,s)\leq \overline{\gamma}s^{p-2}$ for $1<p<2$, ensures that $\gamma^T(x,s)\geq \gamma^A(x,s)$ for a.e. $x\in\Omega$ and $s>0$, when $T\subseteq A$, i.e. that
$    T\subseteq A \Longrightarrow \overline{\Lambda}^T\geqslant\overline{\Lambda}^A,$
giving \Cref{doppia_implicazione_casop}.
\end{proof}

\begin{remark}
    The bounds $\gamma_a(x,s)\geq \underline{\gamma}\left(\frac{s}{s_0}\right)^{p-2}$ in case (2) and $\gamma_a(x,s)\leq \overline{\gamma}\left(\frac{s}{s_0}\right)^{p-2}$ in case (3) are necessary for the Monotonicity Principle (MP) to hold. Indeed, without these additional conditions, the material properties of the background and the anomalous region intersect, preventing a pointwise ordering between them. This, in turn, precludes constructing a material property for a test anomaly $T$ that guarantees $T\subseteq A \Longrightarrow \overline{\Lambda}^T\leqslant\overline{\Lambda}^A$ for case (2) (respectively $T\subseteq A \Longrightarrow \overline{\Lambda}^T\geqslant\overline{\Lambda}^A$ for case (3)) while simultaneously coinciding with $\gamma_b$ in $\Omega\setminus T$.
\end{remark}

\begin{remark}
Theorem \ref{thm_converse_p_esteso} complements the existing theory developed for the $p$-Laplacian variant of the Calderón problem (see  \cite{Salo2012_IP,brander2015enclosure,brander2016calderon,brander2018superconductive,guo2016inverse,brander2018monotonicity,hauer2015p}). In particular, it relaxes the hypothesis of a $p$-Laplacian model for the material filling the anomalous region, allowing for a general quasilinear constitutive relationship, although the growth exponent $q$ must be the same as the background material. However, it is worth noticing that, when $q\neq p$, no order relation between the corresponding material properties can be established in general. As a consequence, the Monotonicity Principle cannot be obtained, and no monotonicity-based result can be obtained in this case.
\end{remark}

\subsection{Quasilinear case}\label{pnot2_quasi}
\begin{figure}[htb]
    \centering
    \includegraphics[width=0.75\linewidth]{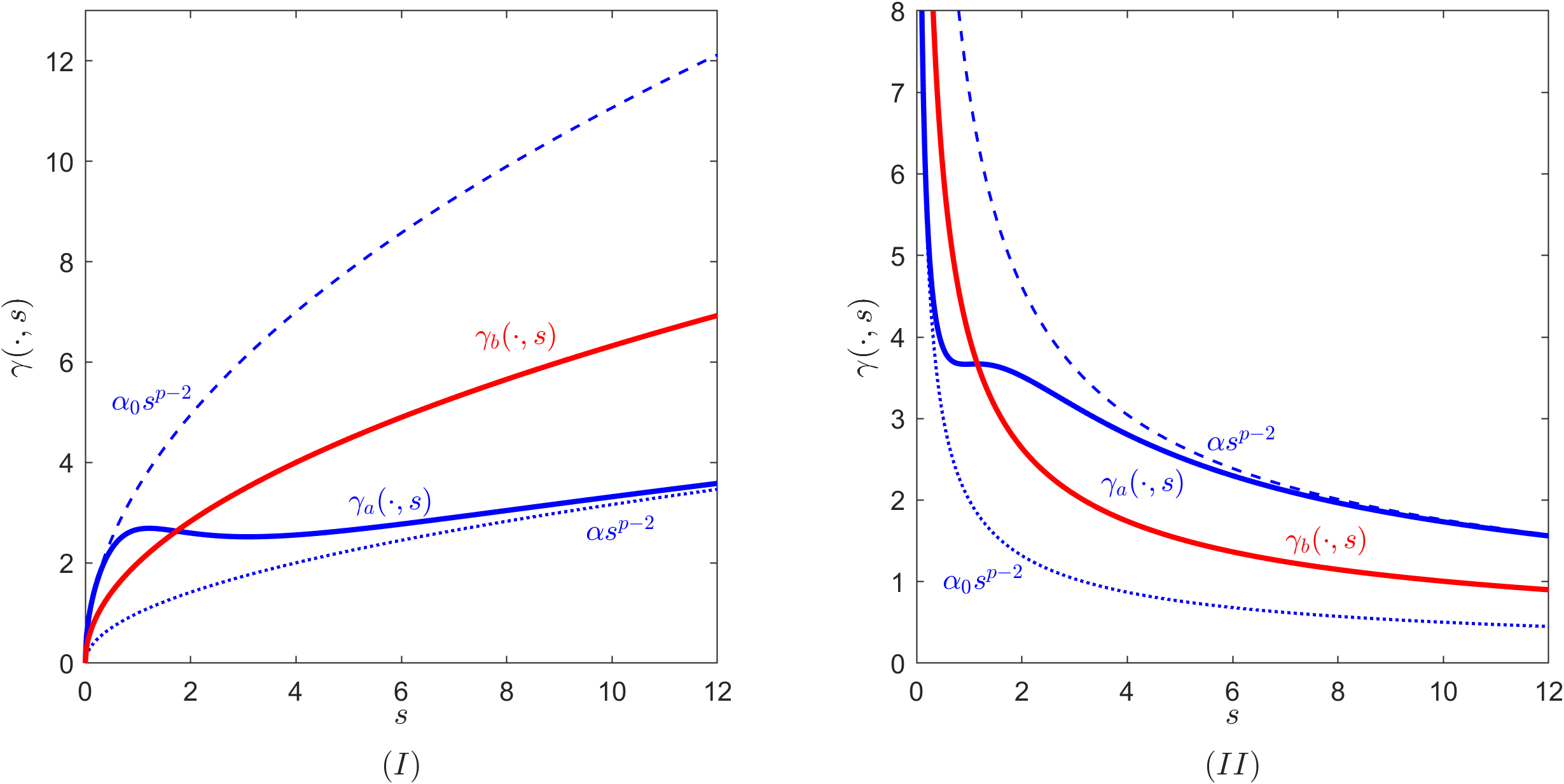}
    \caption{Examples of material properties fulfilling the hypothesis of \Cref{thm_np_1} and \Cref{thm_np_2}.}
    \label{fig_06_quasi_pn2}
\end{figure}

This section is devoted to solving the inverse obstacle problem in the presence of quasilinear materials for both the background and the anomalous region. Unlike in \Cref{sec_p=2_quasi}, the background material property is not required to be bounded, meaning $p\neq 2$ (see \Cref{fig_06_quasi_pn2} for illustrative examples).

More specifically, \Cref{thm_np_1,thm_np_2} show that the result of \Cref{thm_converse_p_esteso} extends beyond $p$-Laplacian backgrounds. Namely, considering a quasilinear, potentially unbounded or vanishing background material property and a quasilinear anomaly, the A-MPM yields a reconstruction contained within the convex hull of the actual anomaly, provided that $p=q$ and that the material properties satisfy the asymptotic behaviour assumptions in either the small or large boundary data regime.

\begin{theorem}[Large boundary data]
\label{thm_np_1}
Let $1<p=q<+\infty$, $p\neq 2$, $f\in X^{p}_\diamond(\partial \Omega)$, $\gamma^A$ and $\gamma^T$ satisfying {\bf (A)}, {\bf (P1)}, {\bf (P2)}, {\bf (Q1)}, {\bf (Q2)}, {\bf (P$_\infty$3)} and {\bf (Q$_\infty$3)}
and one of the following holds
\begin{enumerate}
\item[(I)] $\alpha(x)>\beta(x)$ a.e. in $x\in\Omega$, and $\beta(x)=\beta$ a.e. $x\in\Omega$.
\item[(II)] $\alpha(x)<\beta(x)$ a.e. in $x\in\Omega$, and $\beta(x)=\beta$ a.e. $x\in\Omega$.
\end{enumerate}
Then, in the case (I) (respectively (II)) it holds
\begin{equation}
\label{converse_negqp}
T \nsubseteq\mathrm{conv}( A) \Longrightarrow \overline{\Lambda}^T\nleqslant\overline{\Lambda}^A \  (\text{respectively }   \overline{\Lambda}^T\ngeqslant\overline{\Lambda}^A)\qquad \forall A, T\in\mathcal S (\Omega).
\end{equation}
Moreover, if $A \in \mathcal S (\Omega)$ is convex, then
\begin{equation}
\label{doppia_implicazioneqp}
T\subseteq A \iff \overline\Lambda^T\leqslant{\overline\Lambda}^A \ (\text{respectively } \overline\Lambda^T\geqslant{\overline\Lambda}^A) \qquad\forall\ T\in\mathcal S (\Omega).
\end{equation}
\end{theorem}
\begin{proof}
The proof is identical to that of \Cref{thm_inf_conv}, up to the following minor modifications. 
First, the power $2$ appearing at $\lambda$ in \eqref{lim_ADtN_DtN} and \eqref{lim_together} has to be replaced by $p$; meanwhile, the ratio $1/2$ in \eqref{lim_ADtN_DtN} and \eqref{lim_together} has to be replaced by $1/p$.

The claim follows from applying the results from \Cref{thm_converse_p_esteso} to the limiting $p-$Laplace problem.

In particular, the resulting limiting DtN operator $\Lambda_{\infty}^A$ arises from the material property
\begin{equation*}
    \gamma^A_{\infty}(x,s)=\begin{cases}
        \alpha(x)s^{p-2} & \text{in } A, \\
        \beta s^{p-2} & \text{in } \Omega\setminus A,
    \end{cases}
\end{equation*}
which, regarding case {\it (I)}, yields to case {\it (1)} of \Cref{thm_converse_p_esteso} when $p>2$, and case {\it (2)} when $1<p<2$.

Analogously, for the case {\it (II)}, it yields to the corresponding cases are {\it (3)} and {\it (4)} of \Cref{thm_converse_p_esteso}, for $p>2$ and $1<p<2$, respectively.
\end{proof}

If the limiting behaviours (see {\bf (P$_0$3)} and {\bf (Q$_0$3)}) for the material properties are satisfied in the limit of small boundary potentials, the following analogous result can be stated.
\begin{theorem}[Small boundary data]\label{thm_np_2}
Let $1<p_0=q_0<+\infty$ be such that $p_0\leq p$, $q_0\leq q$, $p_0\neq 2$, $f\in X^{p}_\diamond(\partial \Omega)$, $\gamma^A$ and $\gamma^T$ satisfying {\bf (A)}, {\bf (P1)}, {\bf (P2)}, {\bf (P$_0$3)}, {\bf (Q1)}, {\bf (Q2)}, {\bf (Q$_0$3)}, {\bf (Q$_0$4)} and {\bf (Q$_0$5)}, and one of the following holds
\begin{enumerate}
\item[(I)] $\alpha_0(x)>\beta_0(x)$ a.e. in $x\in\Omega$, and $\beta_0(x)=\beta_0$ a.e. $x\in\Omega$.
\item[(II)]  $\alpha_0(x)<\beta_0(x)$ a.e. in $x\in\Omega$, and $\beta_0(x)=\beta_0$ a.e. $x\in\Omega$.
\end{enumerate}
Then, in the case (I) (respectively (II) it holds
\begin{equation}
T \nsubseteq\mathrm{conv}( A) \Longrightarrow \overline{\Lambda}^T\nleqslant\overline{\Lambda}^A \  (\text{respectively }   \overline{\Lambda}^T\ngeqslant\overline{\Lambda}^A)\qquad \forall A, T\in\mathcal S (\Omega).
\end{equation}
Moreover, if $A \in \mathcal S (\Omega)$ is convex, then
\begin{equation}
T\subseteq A \iff \overline\Lambda^T\leqslant{\overline\Lambda}^A \ (\text{respectively } \overline\Lambda^T\geqslant{\overline\Lambda}^A) \qquad\forall\ T\in\mathcal S (\Omega).
\end{equation}
\end{theorem}

\section{Conclusions}
\label{sec_conclusions}

In this work, a generalisation of the A-MPM for solving the inverse obstacle problem for new classes of nonlinearity is proposed.

First, it is proved that the A-MPM guaranties reconstruction of the anomaly's convex hull for a $p$-Laplace background when $q=p$, even if the anomaly itself is not $q$-Laplace. 
Second, it is demonstrated that the $p$-Laplace Signature (pLS), combined with the A-MPM, enables the reconstruction of the outer support of $A$ when the background is bounded ($p=2$) and $q=p$, or its convex hull when $p \neq 2$ and $q=p$. Moreover, it allows to apply A-MPM when the problem is only AWS and not WS. 

The proposed approach extends A-MPM to configurations in which arbitrary quasilinear materials are present both in the anomaly and in the background, under the constraint of either $p=2$ and arbitrary $q$ or $p \ne 2$ and $q=p$, leaving aside only the case $p \ne q$ and $q \ne p$.

The case $p \ne 2$ and $q \ne p$ will be part of a future work, as well as a numerical implementation and validation to confirm the effectiveness of the proposed strategies, in terms of quality of the reconstructions and robustness to the measurement noise.

\section*{Acknowledgments}
This work has been partially supported by Horizon Europe - COST (project CA24122) and by GNAMPA of INdAM.

\section*{Authorship contribution statement}

{\bf A. Tamburrino}: Conceptualization, Methodology, Formal analysis, Writing, Supervision.

{\bf V. Mottola, G. Piscitelli}: Conceptualization, Methodology, Formal analysis, Writing. 

\bibliographystyle{
siam}

\bibliography{bibliography}

\end{document}